\documentclass[a4paper,11pt]{article}
\pdfoutput=1 

\usepackage{amsmath,amssymb,graphicx}
\usepackage{jcappub}
\usepackage{epsf}
\usepackage{graphicx,epsfig}
\usepackage{amsfonts}
\usepackage{amssymb}
\usepackage[normalem]{ulem}
\usepackage{setspace}
\usepackage{caption}
\usepackage{epstopdf}
\usepackage{color}
\newcommand{\nc}{\newcommand}
\nc{\ba}{\begin{eqnarray}}
\nc{\ea}{\end{eqnarray}}
\newcommand\be{\begin{equation}}
\newcommand\ee{\end{equation}}

\newcommand{\calP}{{\cal{P}}}

\newcommand{\calC}{{\cal{C}}}
\newcommand{\calD}{{\cal{D}}}

\nc{\x}{{\bf{x}}}
\nc{\bfk}{{\bf{k}}}

\def\high{\vphantom{\Biggl(}\displaystyle}

\newcommand{\bea}{\begin{eqnarray}}
\newcommand{\eea}{\end{eqnarray}}
\newcommand{\barr}{\begin{array}}
\newcommand{\earr}{\end{array}}
\def\bal#1\eal{\begin{align}#1\end{align}}

\title{\boldmath The Future of Primordial Features with Large-Scale Structure Surveys}
\author[a,b]{Xingang Chen,}
\author[c]{Cora Dvorkin,}
\author[d]{Zhiqi Huang,}
\author[a,b]{Mohammad Hossein Namjoo,}
\author[e,f,g]{Licia Verde}
\affiliation[a]{ Institute for Theory and Computation, Harvard-Smithsonian Center for Astrophysics, 60 Garden Street, Cambridge, MA 02138, USA}
\affiliation[b]{\center Department of Physics, The University of Texas at Dallas, Richardson, TX 75083, USA}
\affiliation[c]{Department of Physics, Harvard University, Cambridge, MA 02138, USA}
\affiliation[d]{School of Physics and Astronomy, Sun Yat-Sen University
,135 Xingang Xi Road, Guangzhou, CHINA, 510275}
 \affiliation[e]{ICREA \& ICC-UB, University of Barcelona (IEEC-UB), Marti i Franques, 1, Barcelona 08028,
Spain}
 \affiliation[f]{Radcliffe Institute for Advanced Study, Harvard University, MA 02138, USA}
\affiliation[g]{Institute of Theoretical Astrophysics, University of Oslo, Oslo 0315, Norway}

\emailAdd{xingang.chen@cfa.harvard.edu}
\emailAdd{dvorkin@physics.harvard.edu}
\emailAdd{huangzhq25@sysu.edu.cn}
\emailAdd{mohammad.namjoo@cfa.harvard.edu}
\emailAdd{liciaverde@icc.ub.edu}

\abstract{Primordial features are one of the most important extensions of the Standard Model of cosmology, providing a wealth of information on the primordial Universe, ranging from discrimination between inflation and alternative scenarios, new particle detection, to fine structures in the inflationary potential. We study the prospects of future large-scale structure (LSS) surveys on the detection and constraints of these features. We classify primordial feature models into several classes, and for each class we present a simple template of power spectrum that encodes the essential physics. We study how well the most ambitious LSS surveys proposed to date, including both spectroscopic and photometric surveys, will be able to improve the constraints with respect to the current Planck data. We find that these LSS surveys will significantly improve the experimental sensitivity on features signals that are oscillatory in scales, due to the 3D information. For a broad range of models, these surveys will be able to reduce the errors of the amplitudes of the features by a factor of 5 or more, including several interesting candidates identified in the recent Planck data. Therefore, LSS surveys offer an impressive opportunity for primordial feature discovery in the next decade or two. We also compare the advantages of both types of surveys.}

\begin{document}

\maketitle

\flushbottom

\section{Introduction}
\setcounter{equation}{0}

Recent results from Cosmic Microwave Background
(CMB) experiments, in particular the Planck space mission \cite{Ade:2011ah,Ade:2015xua}, have confirmed to high confidence a Standard Model of cosmology in which the
Universe is  spatially flat, with  nearly Gaussian primordial perturbations with a power-law power spectrum which is, close to but not, scale-invariant.
These characteristics, together with the observation of super-horizon perturbations via the large-scale temperature-E-mode polarization anti-correlation \cite{Peiris:2003ff}, are all consistent with predictions of the simplest inflationary models \cite{Guth:1980zm,Linde:1981mu,Albrecht:1982wi}, and make possible to distinguish among different inflationary models, and between inflation and alternative-to-inflation models (e.g.~\cite{Ade:2015lrj, Martin:2013tda}).
In the simplest single-field models of inflation,  the inflaton field driving inflation is a canonical scalar field slowly rolling on a smooth potential.
While classes of inflationary models and alternative-to-inflation models have been ruled out by  available data, future data will make possible to address the next challenges:  to seek further evidence for inflation, and to shed light on the physics of inflationary models. 
There are three  complementary routes: primordial non-Gaussianity, which probes the (self)interactions of fields in the primordial Universe; primordial gravity waves, as they leave a signature in the CMB polarization pattern and are directly related to the energy scale of the primordial Universe; and departures from a smooth power spectrum in the form of specific features.
In this paper we study the imprints of the latter signals on the large-scale structure of the Universe and make forecasts of their detectability.

An important class of beyond-Standard-Model feature signals is the scale-dependent oscillatory signal in the power spectrum of the curvature perturbations. Depending on their natures, these signals provide a variety of valuable information on the physics of the primordial Universe. Broadly speaking, we will study the following several classes of models, their characteristic signals and the associated underlying physics: 1) The sharp feature models, in which the feature signal in the power spectrum oscillates linearly in scale, indicates that there is a sharp feature in the model that temporarily breaks the slow-roll condition, or equivalently that there is injection of new physics at a certain time for all scales; 2) The resonance models, in which the feature signal oscillates logarithmically in scale, indicates that there are periodic features in the model, or equivalently that there is injection of new physics at a certain scale for all time; 3) Classical standard clock models, in which the feature signal takes a special combination of the sharp and resonance form, indicates the excitation of massive fields that subsequently oscillate. The first two classes of models are studied under the assumption of the inflationary scenario, and provide details of inflation models. The third class of models is studied without assuming inflation, and the oscillation of the massive field is used as the standard clock to record the scale factor evolution of the background $a(t)$, hence providing a distinction between the inflation and alternative-to-inflation scenarios. Besides oscillatory signals, 4) we also consider a type of feature models of which the signal is mostly a bump in the momentum space; such a signal typically indicates large interactions among fields at a certain moment.
Therefore, discovering features in the primordial density perturbations would have major impacts in the field of cosmology: probing the nature of inflation models; unveiling direct clues that distinguish between inflation and alternative scenarios; and discovering new and heaviest particles ever found.
These models will be reviewed and classified in Sec.~\ref{sec:featuremodels}. Also see Refs. \cite{Chen:2010xka,Chluba:2015bqa} for more detailed reviews on some of the models.

Besides the CMB, deviations from a featureless primordial power spectrum can be probed by large-scale structure (LSS) surveys.
The golden era  of mapping  the LSS of the Universe is still to come. The near-future galaxy surveys can be divided into two types that are complementary to each other: spectroscopic and photometric. Here we consider both.

A spectroscopic Euclid-like wide survey \cite{Amiaux:2012bt,Laureijs:2011gra,Amendola:2012ys} will measure about $10^7$ galaxies up to $z\sim 2$. The advantage of a spectroscopic survey is that it can obtain much better resolution in radial direction. To break the degeneracy between the amplitude of primordial power spectrum and galaxy biases, it is necessary to use redshift distortion information (in combination with CMB data), which requires good redshift resolution.
Spectroscopic surveys conserve  -- to a good approximation -- the full  three-dimensional information of the matter power spectrum and maximize the number of independent Fourier modes that can be sampled in a given volume.

A photometric LSST-like survey \cite{LSST} will provide photometry for $10^9$-$10^{10}$ galaxies, from the local group up to high redshift $z\gtrsim 5$ and include very faint samples (magnitude $\sim 28$). The high-redshift galaxy samples provide valuable information about galaxy formation and galaxy evolution. For our purpose of studying primordial signals, it is sufficient to consider galaxies samples up to $z\sim 3$, where the galaxy detection rate is still high enough to guarantee a sufficiently low shot noise.
While spectroscopy is required to measure the redshift of a tracer (typically a galaxy) to high precision, a redshift estimate  can also be obtained, usually more rapidly and cheaply, with broad-band photometry. Photometric surveys can cover a large area of the sky faster than a spectroscopic survey and reach a higher number density of observed objects. The drawback is that some of the three-dimensional information present in spectroscopic surveys is lost due to the more uncertain, photometric,  redshift determination.

For feature models, there may be a trade-off between volume and number density  sampled (optimized by photometric surveys) and  full 3-D information (optimized by spectroscopy surveys).  This is what we want to  explore, presenting  forecasts for the most ambitious photometric ground-based survey  (LSST) and spectroscopic space-based mission (Euclid) proposed to date.

Forecasts for future spectroscopic Euclid-like survey on feature models were made in Ref.~\cite{Huang:2012mr}, using two specific feature models. As highlighted in \cite{Huang:2012mr}, such a survey will be crucial to detect and measure features in the inflationary potential. In the CMB what is observed ultimately is a two-dimensional projection of the three dimensional power spectrum; features are smoothed out by this projection. LSS preserves (at least in part) the three-dimensional information in the sky and, moreover, the number of (Fourier) modes probed by future surveys will be much larger than the number of modes measured by the Planck satellite. In this case, the limitation on the width of the detectable feature is given by the volume of the survey $\lambda_{\rm min} \sim V^{1/3}$. Despite non-linearities limiting the information that can be extracted, Ref.~\cite{Huang:2012mr} shows that LSS observations will quantitatively and qualitatively improve constraints on primordial features.

Here we include a broader class of models and benchmark points, some of which are motivated by various best-fit (although statistically insignificant at present) models in comparison with the recent Planck data. Furthermore, to efficiently search for the broad scope of physics and to have a big picture of what to expect from future experiments, we decide to use a model-independent approach at the cost of losing certain model-dependent details. Instead of studying specific models, we use simple templates to encode entire classes of models. Such templates carry the signature of specific physical processes in each class described above (see also Sec.~\ref{sec:featuremodels}).  This way, our analysis is  sensitive to essential physics behind the models rather than to model-specific parameters. For forecasts, this approach gives us an order-of-magnitude estimate of how well each class of models will be constrained by future experiments; for future data-analyses, this approach may be applied at an initial stage and followed up by more model-dependent analysis if any positive signal is found.

We will consider both the Euclid-like spectroscopic survey and LSST-like photometric survey, in combination with a CMB Planck prior, and compare the results between each other, and between them and the current Planck-like constraints. We study how much LSS surveys will advance our knowledge in primordial feature models in comparison with the current Planck 2015 constraints, and address the advantages of both types of surveys.

The rest of the paper is organized as follows. In Sec.~\ref{sec:featuremodels} we review and classify the feature models we consider in this work, and we introduce templates for each class of models.  The adopted methodology is described in Sec.~\ref{sec:Fisher}. The results are presented in Sec.~\ref{sec:results}, and we conclude in Sec. \ref{sec:conclusions}.

\section{Features in primordial density perturbations}
\label{sec:featuremodels}
In this section, we classify primordial feature models into several classes and provide a brief review for each of them.
For each class of model, we introduce a simple analytic template which captures the most important effects of the models in the matter power spectrum.
These simple templates will be used in Sec.~\ref{sec:Fisher} to forecast the detectability of feature models in future large-scale structure surveys.
Of course, in each class there are also many model-dependent details that are not captured in our simple templates, but in this paper we concentrate on the leading properties that tell us the qualitative big picture.
If these leading properties were to be discovered by future experiments, more detailed model-dependent analyses should then follow.

It is easiest to present the templates in terms of the correction to the power spectrum of the curvature perturbation, i.e. $\Delta P_\zeta /P_{\zeta0}$.  Here $\Delta P_\zeta$ encodes the feature properties and
\ba
P_{\zeta0} = A_s \left(\dfrac{k}{k_0} \right)^{n_s-1},
\ea
which is the power spectrum for a smooth $\Lambda$CDM model, and $A_s$, $n_s$ and $k_0$ are the amplitude, the scalar spectral index and the pivot scale, respectively.

\subsection{Sharp feature signal}
\label{Sec:Sharp}

Sharp features are localized features in inflationary potentials or internal field space that temporarily break the slow-roll condition. This excites the quantum fluctuations of the curvature mode near and inside the horizon, and generates a special type of scale-dependent oscillating feature in density perturbations \cite{Starobinsky:1992ts}, which we call the ``sharp feature signal".

In terms of model building, the nature of sharp features can vary a lot. For example, sharp features can be a kink, step or bump in the single-field inflationary potential \cite{Starobinsky:1992ts,Adams:2001vc,Chen:2006xjb,Mortonson:2009qv,Dvorkin:2009ne, Adshead:2011jq,Adshead:2011bw,Hazra:2014goa,Bartolo:2013exa}; such features can also appear in the internal field space such as the sound speed of the inflaton \cite{Bean:2008na,Miranda:2012rm,Bartolo:2013exa}, or appear in potentials in the multi-field space \cite{Chen:2014joa,Chen:2014cwa}; the sharp feature can also be a sharp bending of the inflaton trajectory in the multi-field space \cite{Achucarro:2010da,Gao:2012uq}; and so on. In density perturbations, despite of many model-dependent details, the signals generated by all these features share a common property, namely a correction with sinusoidal wiggles as a function of the scale (i.e.,~sinusoidal running). This running behavior has highly correlated signals in both the power spectrum and the three-point function \cite{Chen:2006xjb,Chen:2008wn,Gong:2014spa,Mooij:2015cxa,Appleby:2015bpw,Romano:2014kla,Palma:2014hra}.

In this work, we study the large-scale structure power spectrum and use the following simple template for the sharp feature signal \cite{Chen:2006xjb,Chen:2011zf,Fergusson:2014hya,Fergusson:2014tza}:
\begin{align}
  \frac{\Delta P_\zeta}{P_{\zeta0}} =
    C \sin \left( \frac{2 k}{k_f} + \phi \right).
    \label{Template_Sharp}
\end{align}

This template has three parameters: the amplitude $C$, the oscillation frequency $1/k_f$  and a phase $\phi$.
Note that, in addition to the leading property captured by the above template, the sharp feature signals also have complicated model-dependent details and mostly manifest as a model-dependent envelop modulating the sinusoidal running in Eq. \eqref{Template_Sharp}.
For simplicity and generality, we have neglected this envelop behavior for the following reasons. If the feature is relatively sharp, the sharp feature signal is very extensive in the momentum space, and the scale dependence of the envelop is less important than that of the sinusoidal running. However, if the feature is relatively mild, the envelop may become important since the sharp feature signal decays quickly in momentum space. The template in Eq. \eqref{Template_Sharp} applies to the former case and, as we will see, this is also the case for which the large-scale structure has more constraining power (see Sec.~\ref{Sec:step} for a concrete example). Furthermore, because this envelop behavior is highly dependent on the nature and sharpness of the feature, for a model-independent data analysis, it may be more effective to ignore it and first analyze the common property in Eq. \eqref{Template_Sharp}. If any candidate signals are positively identified in such model-independent data analyses, the envelop behavior may be added and play an important role in subsequently ruling models out and constraining the parameters of viable models.
In this case, we expect to be able to distinguish the sharpness of the feature better than the nature of the feature.

\begin{figure}
\center
\includegraphics[scale=.8]{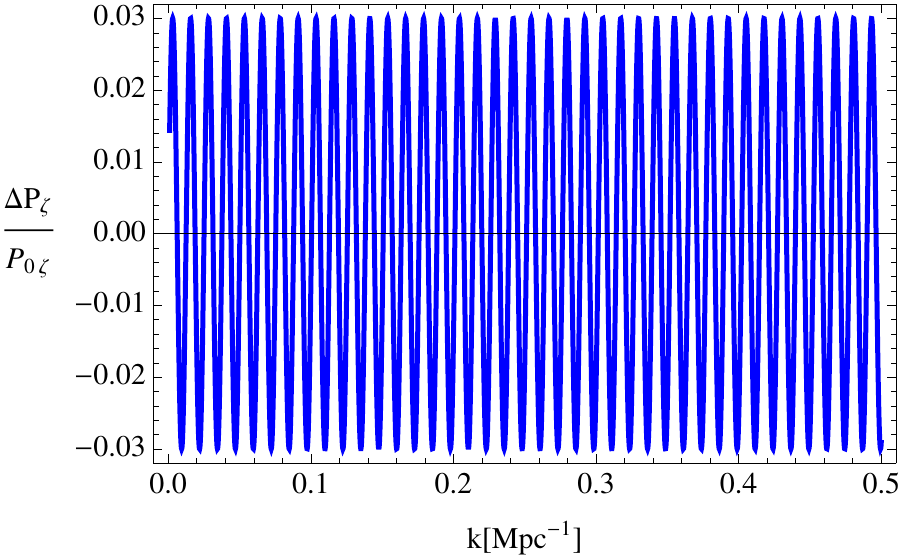}
\hspace{.5cm}
\includegraphics[scale=.8]{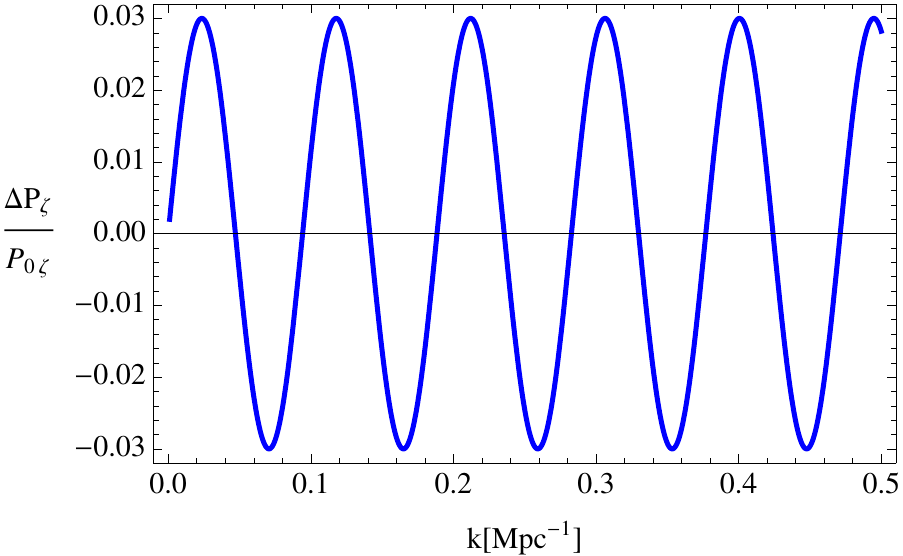}
\hspace{.75cm}
\includegraphics[scale=.8]{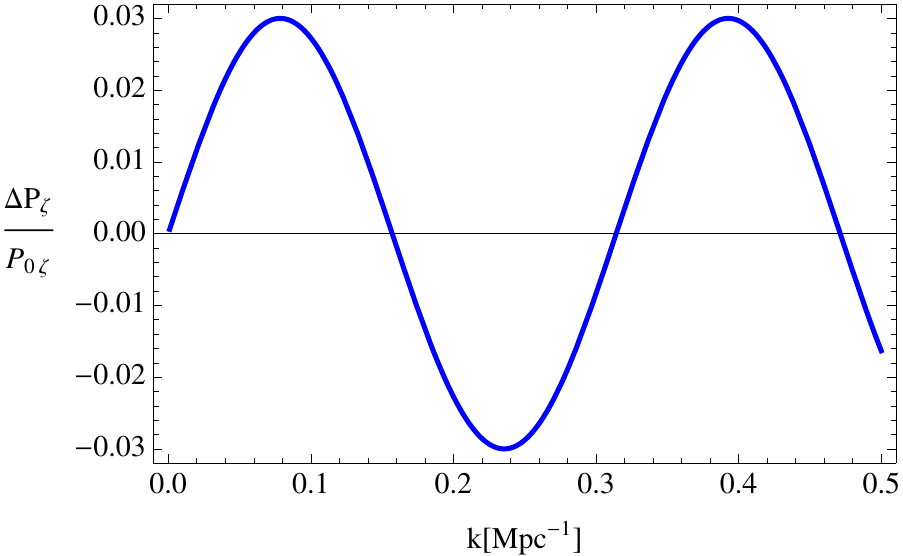}
\hspace{.5cm}
\caption{\small Primordial sharp feature signal. Here we have set, $k_f [{\rm Mpc}^{-1}] = 0.004$ (upper-left), 0.03 (upper-right) and 0.1 (lower). For all plots we also have  $C=0.03$ and $\phi=0$. }
\label{Fig:sin}
\end{figure}

We expect the sensitivity of the LSS to the primordial oscillatory features to highly depend on the oscillatory frequencies; we use the following several frequencies as benchmarks:
$k_f = 0.004$ ${\rm Mpc}^{-1}$, $k_f = 0.03$ ${\rm Mpc}^{-1}$ and $k_f = 0.1$ ${\rm Mpc}^{-1}$. We also choose the amplitude to be $C=0.03$, which is small enough to be consistent with Planck constraints \cite{Ade:2015lrj} (see Fig.~\ref{Fig:sin} where we have plotted the template for each frequency).

The sharp features are compared with Planck data in \cite{Ade:2015lrj,Hu:2014hra,Chen:2014cwa,Hazra:2016fkm,Hamann:2007pa,Benetti:2011rp,Dvorkin:2011ui,Hazra:2013xva,Meerburg:2013dla,Meerburg:2013cla,Benetti:2013cja,Meerburg:2014bpa,Meerburg:2014kna,Planck:2013jfk}. These analyses show that there is an interesting sharp feature candidate at around $\ell \sim 20-40$ (with $k_f\simeq 0.004/$Mpc \cite{Adshead:2011jq,Benetti:2011rp,Chen:2014cwa}) with marginal statistical significance \cite{Ade:2015lrj}, which is also seen in the WMAP data at similar significance \cite{Hinshaw:2003ex}. We discuss this case in Sec.~\ref{Sec:step}. There is also another sharp feature candidate with $k_f\sim 0.0038/{\rm Mpc}$ that mainly picks up signal around $\ell \sim 700-800$ \cite{Chen:2014cwa,Hu:2014hra}. Other than these, the relative amplitude of the sharp feature signals $C$ has been constrained to be below a few percent, and details vary depending on the frequency.

\subsubsection{A concrete example: a step in the potential}
\label{Sec:step}
\begin{figure}
\center
\includegraphics[scale=.8]{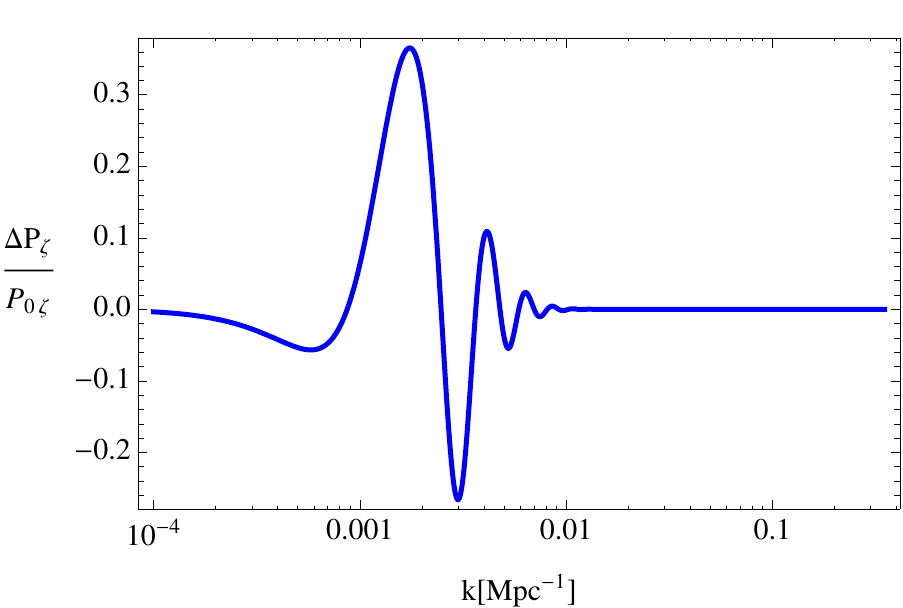}
\includegraphics[scale=.8]{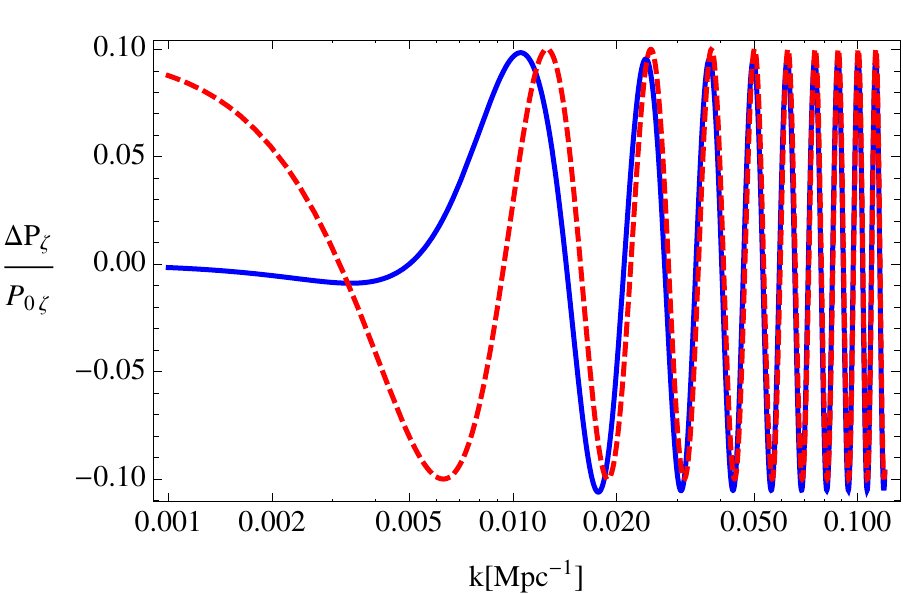}
\caption{\small
Two examples of the sharp feature signals from the step-in-potential model (Eq. \eqref{Template:step}). The {\em left panel} is a milder step (i.e.~the step has relatively small height-to-width-ratio) with $\calC = 0.218$, $\eta_f = 1.44$ Gpc and $x_d=1.6$. As we can see, the envelop of the sinusoidal running is very important for this case.
The solid-blue line in the {\em right panel} is a sharper step (i.e.~the step has relatively large height-to-width-ratio) with $\calC = 0.033$, $\eta_f=0.25$ Gpc and $x_d=100$. As we can see, the envelop of the sinusoidal running is less important, and the simple template of Eq. \eqref{Template_Sharp} (dashed-red line, $C=0.1$, $k_f=0.004/$ Mpc and $\phi=\pi/2$) is a good approximation in this case.}
\label{Fig:step}
\end{figure}

As mentioned earlier, besides the sinusoidal running, there is typically a scale-dependent envelop which differs among models. Here we study one specific model, which has the advantage that there are already best fit values obtained by CMB data analysis, but has the disadvantage that the envelop for the best fit case is so strong that almost erases the signal at small scales where significant improvement is expected from LSS data.  This will therefore help us assess how much the signal-to-noise may be overestimated  by our analysis that  uses the unsuppressed  template  of Eq.~\eqref{Template_Sharp}. This model has been proposed to explain the glitches in CMB data at scales corresponding to $\ell \sim 20-40$ \cite{Mortonson:2009qv,Miranda:2014fwa}.
The template, first proposed in Ref. \cite{Adshead:2011jq}, is
\ba
\dfrac{\Delta P_\zeta}{P_{\zeta0}} = \exp[\calC \times \calD  (k\eta_f ) ]-1,
\label{Template:step}
\ea
in which $\calC$ is the amplitude of the feature, $\eta_f$ corresponds to the conformal time of the phase transition during inflation and
\ba
\calD  (x) = \dfrac{(x/x_d)}{\sinh(x/x_d)} \times
\left[  \left( -3+\dfrac{9}{x^2} \right) \cos(2x)
+\left( 15- \dfrac{9}{x^2}  \right) \dfrac{\sin(2x)}{2x}
\right] .
\ea
The best fit values for the parameters are found to be   $\eta_f \simeq 1.44$ Gpc, $\calC=0.218$ and $x_d=1.6$ \cite{Adshead:2011jq}.
This feature can be generated by a step in the potential of the inflaton field. We plot this template with the above parameters in Fig.~\ref{Fig:step} ({\em left panel}) from which it is clear that the statistics for this model and this choice of parameters will likely not improve much with LSS observations since the signal decays away rapidly at small scales.

The significant difference between this example and the sharp feature template is due to the specific parameters we used to fit the $\ell\sim 20-40$ glitch. Namely, it is a rather mild step, generating very few ringings in momentum space. For a sharper step, it can be shown that for small $\calC$ and large $x_d$ the template for the step model in Eq.~\eqref{Template:step} reduces to $3 \calC \cos(2 k \eta_f)$  in the large $k\eta_f$ limit, which is consistent with the simple template in Eq.\eqref{Template_Sharp}. See the right panel of Fig. \ref{Fig:step} for a comparison of two templates in this limit. Note that large $x_d$ corresponds to a very sharp (very short timescale) step.
This comparison illustrates the points emphasized  in the discussion below Eq.~\eqref{Template_Sharp}.

\subsection{Resonance feature signal}
\label{Sec:Resonance}

Features in inflationary models that are periodic (or semi-periodic) in time generate another type of feature signals in the density perturbations. The most important property for this type of features is not the sharpness, but the periodicity of the features in the inflationary potential or internal field space.
These features generate a time-dependent and oscillatory component in the background parameters. If the frequency of this oscillation is much larger than the Hubble parameter $H$, the background oscillation will resonate with the quantum fluctuations of the curvature mode inside the horizon, generating another special type of scale-dependent oscillatory signals in density perturbations \cite{Chen:2008wn}, which we call the ``resonance feature signal".

In terms of model-building, the resonance models may be realized in string theory as large-field models such as the axion monodromy inflation \cite{Flauger:2009ab}, or small-field models such as the brane inflation \cite{Bean:2008na}, or in particle physics in terms of axion inflation \cite{Wang:2002hf}. Special types of resonance signals can also arise due to other (semi-)periodic oscillations such as the oscillation of massive fields \cite{Chen:2011zf,Chen:2011tu,Chen:2012ja}, which we review in Sec. \ref{Sec:SC}. Like the sharp feature case, the resonance features also have highly correlated signals between the power spectrum and the bispectrum \cite{Chen:2008wn,Flauger:2010ja,Chen:2010bka}, but we do not consider those correlations here.

\begin{figure}
\center
\includegraphics[scale=.8]{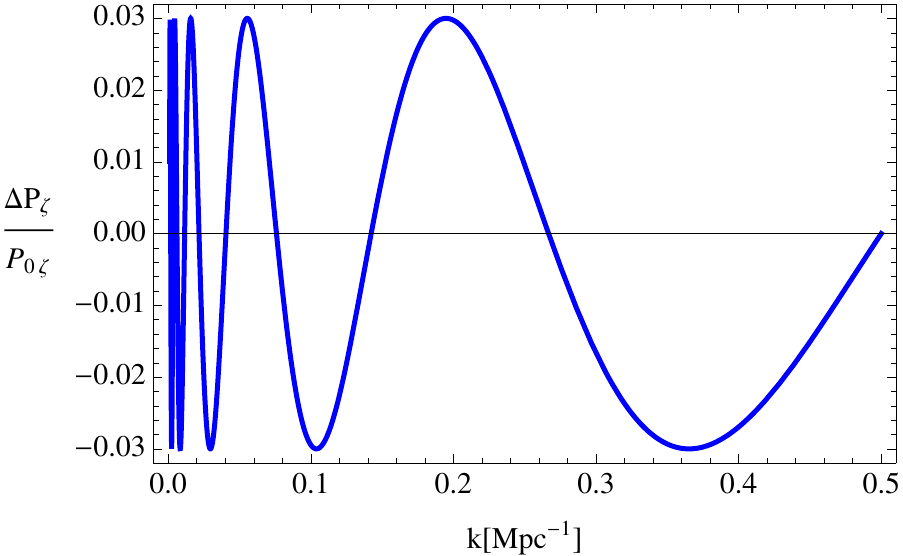}
\hspace{.5cm}
\includegraphics[scale=.8]{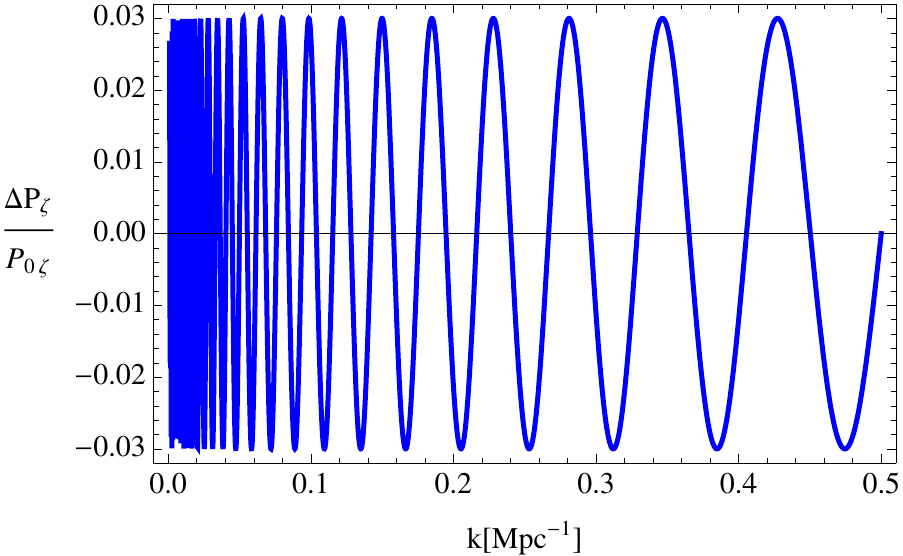}
\hspace{.5cm}
\includegraphics[scale=.8]{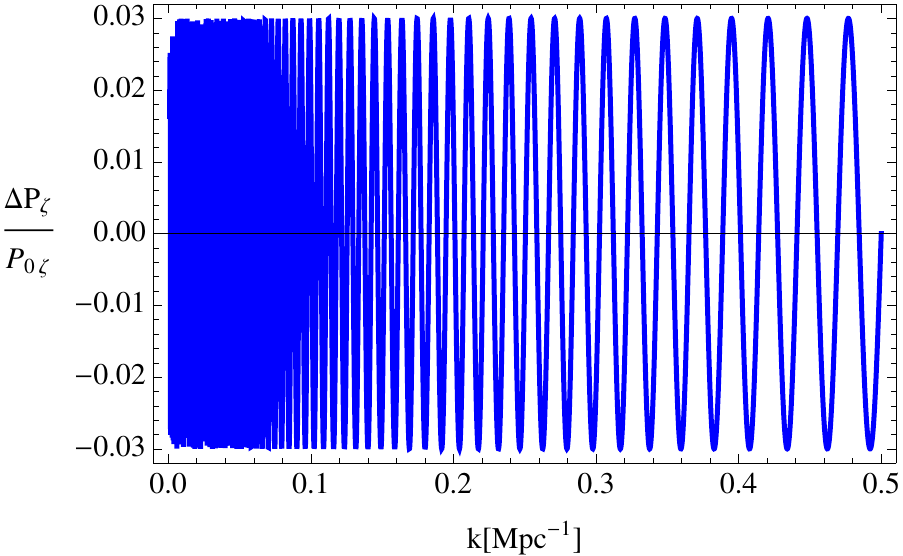}
\caption{\small The resonance feature in the power spectrum, as in Eq. \eqref{Template_Resonance}. For all three figures we have set $C=0.03$ and $\phi=0$, while for the upper-left, upper-right and lower plots we have $\Omega=5, 30$, and $100$, respectively.}
\label{Fig:res}
\end{figure}

The power spectrum template for resonance model is given by \cite{Chen:2008wn}:
\begin{align}
\frac{\Delta P_\zeta}{P_{\zeta0}} = C \sin \left[ \Omega \log \left( 2k \right) + \phi \right] ~.
\label{Template_Resonance}
\end{align}
This template has three parameters: an amplitude $C$, a frequency $\Omega$, and a phase $\phi$. One can add another dimensionful parameter to make the argument of the logarithm dimensionless. However, this parameter is totally degenerate with the phase $\phi$, hence we absorbed it in the phase by a simple redefinition.  This makes the template non-degenerate, thus easier to compare with the data.
Similarly to the sharp feature signal, the expected signal-to-noise should depend on the frequency. For reference, we consider the cases: $\Omega=5$, $\Omega=30$, and $\Omega=100$, and we use $C=0.03$ for the amplitude of the signal (see Fig.~\ref{Fig:res}).

The resonance features are compared with Planck data in \cite{Ade:2015lrj,Flauger:2014ana,Martin:2003sg,Pahud:2008ae,Aich:2011qv,Meerburg:2013dla,Meerburg:2013cla,Meerburg:2014kna,Planck:2013jfk}, where it has been pointed out that
there is an interesting resonance feature candidate with $\Omega\sim 30$ that mainly picks up signal around $\ell \sim 700-800$ \cite{Ade:2015lrj}.
Other than that, the relative amplitude of the resonance signal $C$ has been constrained to be below a few percent with details depending on the frequency of the feature.

The two classes of feature models, namely the sharp feature and resonance feature models, can also be realized in terms of non-Bunch-Davies (non-BD) vacua in inflationary models (see Ref. \cite{Ade:2015lrj} for a summary). In these models, some new-physics scales are introduced hypothetically somewhere inside the horizon; the quantum modes coming out from these scales take non-BD vacuum forms, and then they follow the evolution of the usual low-energy field theory. If the new-physics is introduced for all modes at a specific instant, the signal generated in density perturbations is like the sharp feature signal in Sec.~\ref{Sec:Sharp}; if the new-physics is introduced for each mode as it reaches the same energy scale, the signal generated is like the resonance feature signal in Sec.~\ref{Sec:Resonance}.
As we can see, the concrete models in Sec.~\ref{Sec:Sharp} \& \ref{Sec:Resonance} are special realizations of these hypothetical new physics (see \cite{Chen:2010xka,Chluba:2015bqa} for further details).

\subsection{Primordial standard clock signal}
\label{Sec:SC}

The oscillation of massive fields is standard in arbitrary backgrounds, and can be regarded as a clock that generates standard ticks. These ticks can be used as standard clocks to directly record the time evolution of the background scale factor $a(t)$ of the primordial Universe in density perturbations. Since $a(t)$ is the defining property of the primordial Universe scenario, measuring the clock signal would  enable us to distinguish inflation from other possible alternative scenarios \cite{Chen:2011zf,Chen:2011tu,Chen:2012ja,Chen:2014joa,Chen:2014cwa,Chen:2015lza,Chen:2016cbe}.
There are two types of standard clock models. The classical standard clock models \cite{Chen:2011zf,Chen:2011tu,Chen:2014joa,Chen:2014cwa} rely on classical oscillations of the massive field, while the quantum standard clock models \cite{Chen:2015lza} rely on quantum fluctuations of the massive field. The latter is more general, but observing it requires the observation of primordial non-Gaussianities \cite{Arkani-Hamed:2015bza,Chen:2009we,Chen:2009zp,Sefusatti:2012ye,Norena:2012yi,Gong:2013sma,Emami:2013lma,Jackson:2012fu,Craig:2014rta,Dimastrogiovanni:2015pla,Schmidt:2015xka,Baumann:2011nk,Assassi:2012zq,Noumi:2012vr}.
In this work, we focus on the classical primordial standard clock model \cite{Chen:2011zf,Chen:2011tu,Chen:2012ja,Chen:2014joa,Chen:2014cwa} and study its effects on the power spectrum.

The classical standard clock signals are a class of feature models that involve a special mixture of two types of features.
In these models, a sharp feature excites classical oscillations of a massive field, and there are two types of signals generated in this process.
The first one is the signal generated by any sharp feature that excites the massive field; the nature of this signal is that of the sharp feature type. The second one is generated by the subsequent oscillations of the massive field; the nature of this signal is that of the resonance type.
The two parts are connected smoothly.
The sharp feature part extends to the largest scales. The resonance part extends to the smallest scales and it directly encodes $a(t)$; hence this is the most important part of the signal, and we call it the ``clock signal". We call the entire signal the ``full standard clock signal".

A complete example of the classical standard clock models can be found in Ref. \cite{Chen:2014joa,Chen:2014cwa}, where it is compared with Planck 2013 data. Below we review the template for the clock signal in a general background, as well as the full standard clock signal for a specific inflationary model studied in Ref. \cite{Chen:2014joa,Chen:2014cwa}.

\subsubsection{Clock signal}
\label{Sec:clock}
In a general background, the behavior of the scale factor  can be modeled by $a \sim (t/t_0)^p$, where $t_0$ is a reference time and $p$ is a parameter that determines the type of the primordial Universe scenarios \cite{Chen:2011zf}: requiring that the modes exit the horizon as time passes, one concludes that $|p|>1$ corresponds to fast expansion scenarios, namely inflation;  ${\cal O}(1)\sim p<1$ corresponds to fast contraction scenarios; $0<p\ll 1$ corresponds to slow contraction scenarios; and $-1\ll p<0$ corresponds to slow expansion scenarios. Note that $t$ runs from $0$ to $\infty$ for $p>1$, and from $-\infty$ to $0$ otherwise.

\begin{figure}
\includegraphics[scale=.8]{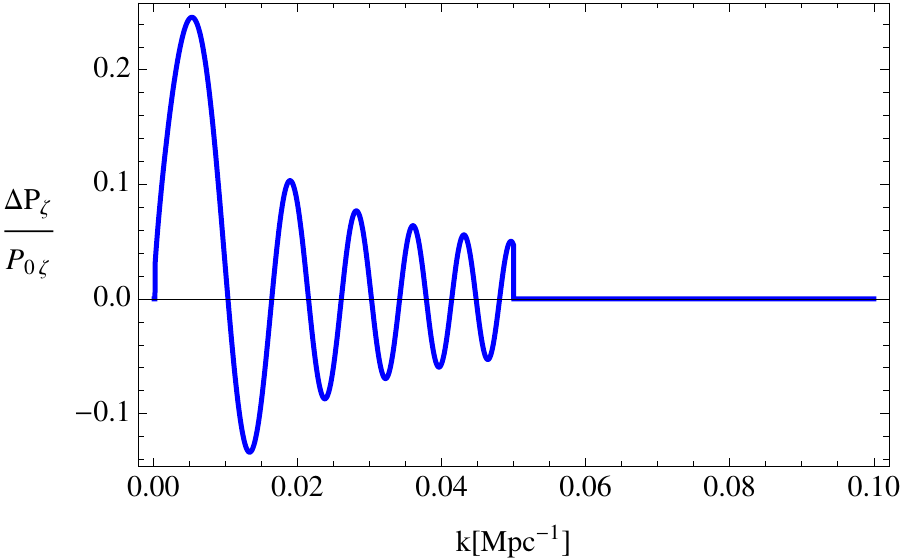}
\includegraphics[scale=.8]{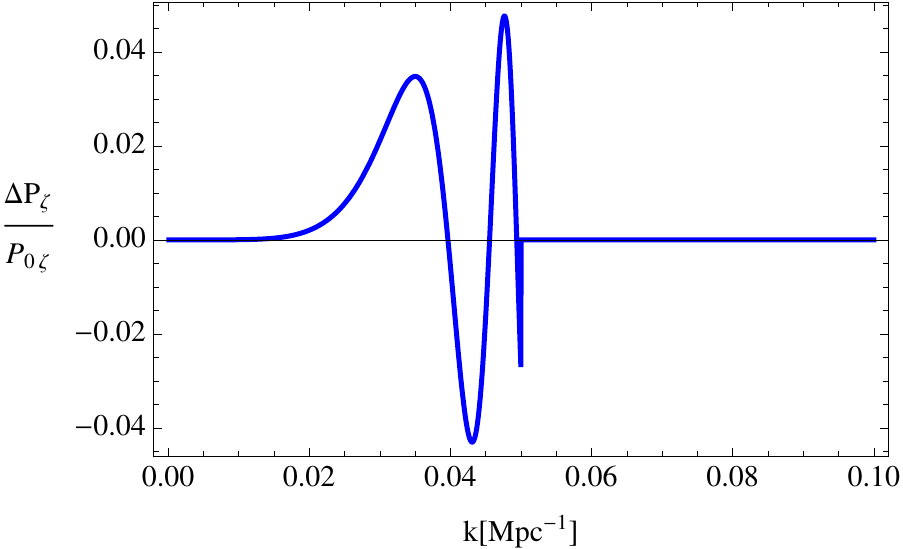}
\center
\includegraphics[scale=.8]{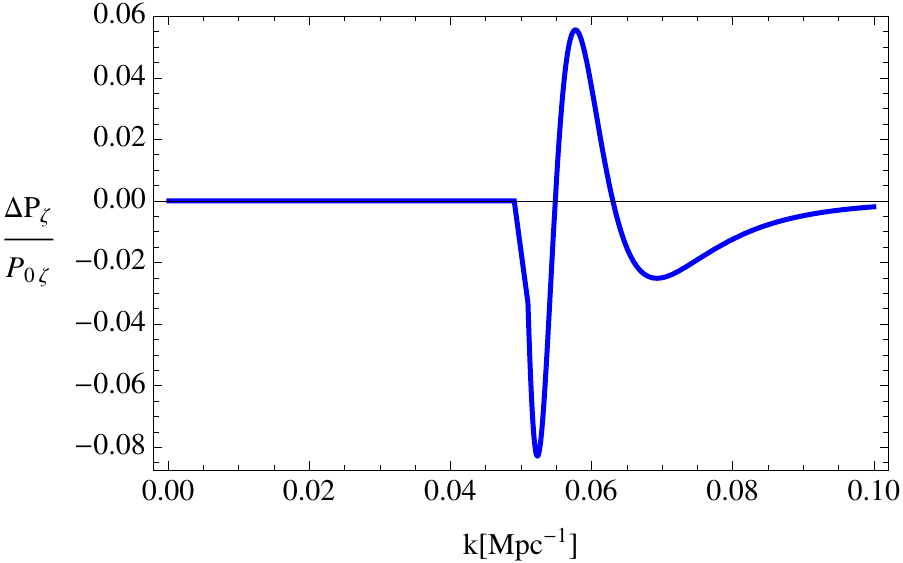}
\hspace{.5cm}
\caption{\small The clock signal for alternative scenarios (Eq.\eqref{Template:clock} with ($-1<p<1$)). We have set $C=0.05$, $k_r=0.1/$Mpc, $\Omega_{\rm eff}= 100$, $\phi=0$, and $p=2/3$ (upper-left panel), $1/5$ (upper-right panel) and $-1/5$ (lower panel).}
\label{Fig:alt_clock}
\end{figure}

The clock signal in a general background can then be described by the following template with five free parameters \cite{Chen:2011zf,Chen:2014cwa},
\begin{align}
  \frac{\Delta P_\zeta}{P_{\zeta0}} = \begin{cases}
    0, & \hspace{.7cm}\text{$k_r > 2 k$ \hspace{.9cm}for expanding scenarios},  \\
    0, & \text{$ \dfrac{2k_r}{ \Omega_{\rm eff}} >2k >  k_r$ \,\,\,\, for contracting  scenarios}, \\
    \high{
    C \left( \frac{2k}{k_r}  \right)^{-\frac{3}{2}+\frac{1}{2p}}
    \sin \left[ \frac{p \Omega_{\rm eff}}{2} \left( \frac{2k}{k_r}  \right)^{\frac{1}{p} } + \phi \right]},  & \text{otherwise},
  \end{cases}
  \label{Template:clock}
\end{align}
where $C$ is the amplitude of the signal, $\phi$ the phase, $p$ determines the time evolution of the scale factor, $\Omega_{\rm eff}$ is the frequency and $k_r$ corresponds to the first mode that resonates with the oscillations of the massive clock field. Note that the mass of the clock field, which is implicit here, controls the value of the frequency $\Omega_{\rm eff}$, as well as the scale $k_r$.
We can see that the evolution of the background scale factor $a(t)$ is encoded as follows: the phase of the clock signal as a function of the scale $k$ is the inverse function of $a(t)$.
We try three different cases: $p=2/3, 1/5$ and $-1/5$, corresponding to the matter contraction (a ``fast contraction" scenario), slow contraction and slow expansion scenarios, respectively. The other parameters are set to: $\phi=0$, $C=0.05$, $\Omega_{\rm eff}=100$, and $k_r=0.1/$Mpc. We show the template for each scenario in Fig. \ref{Fig:alt_clock}.

\begin{figure}
\center
\includegraphics[scale=.8]{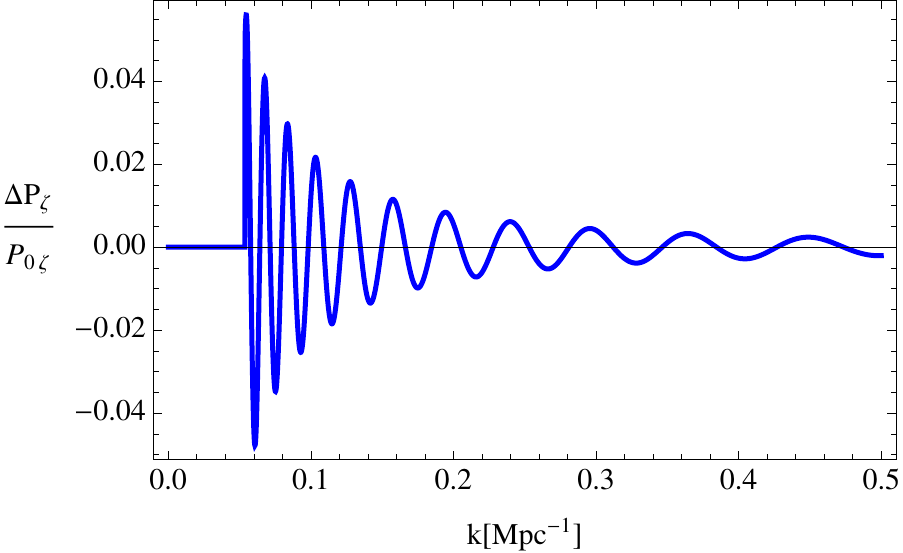}
\hspace{.5cm}
\caption{\small The clock signal  for inflationary scenarios (Eq.\eqref{Template:inf_clock}) with $k_r=0.10753/$Mpc, $p=105$, $\Omega_{\rm eff}= 59.1$, $\phi=2.15$, and $C=0.0576$.}
\label{Fig:inf_clock}
\end{figure}

The other scenario, which has been analyzed in Ref.~\cite{Chen:2011zf,Chen:2014cwa}, is the inflationary scenario where $|p|\gg 1$. Note that  for sufficiently large $p$, the template is effectively independent of $p$. i.e., one can write a simplified template for inflation as follows:
\begin{align}
  \frac{\Delta P_\zeta}{P_{\zeta0}} \xrightarrow{p\gg 1} \begin{cases}
    0, & \hspace{.1cm}\text{$k_r > 2 k$ },  \\
    \high{
    C \left( \frac{2k}{k_r}  \right)^{-\frac{3}{2}}
    \sin \left[ \frac{ \Omega_{\rm eff}}{2} \ln\left( 2k  \right) + \phi \right]},  \qquad
   & \text{otherwise},
  \end{cases}
  \label{Template:inf_clock}
\end{align}
where we have redefined the phase for convenience.
For the fiducial values, we consider the best-fit from Planck data obtained in Ref.~\cite{Chen:2014cwa}; namely we set ($k_r$, $\Omega_{\rm eff}$, $\phi$, C) = ($0.10753$/Mpc, $59.1$, $4.07$, $0.0576$) (see Fig. \ref{Fig:inf_clock}).

\subsubsection{An example of full standard clock signal}
\label{Sec:SC_full}
\begin{figure}
\center
\includegraphics[scale=.8]{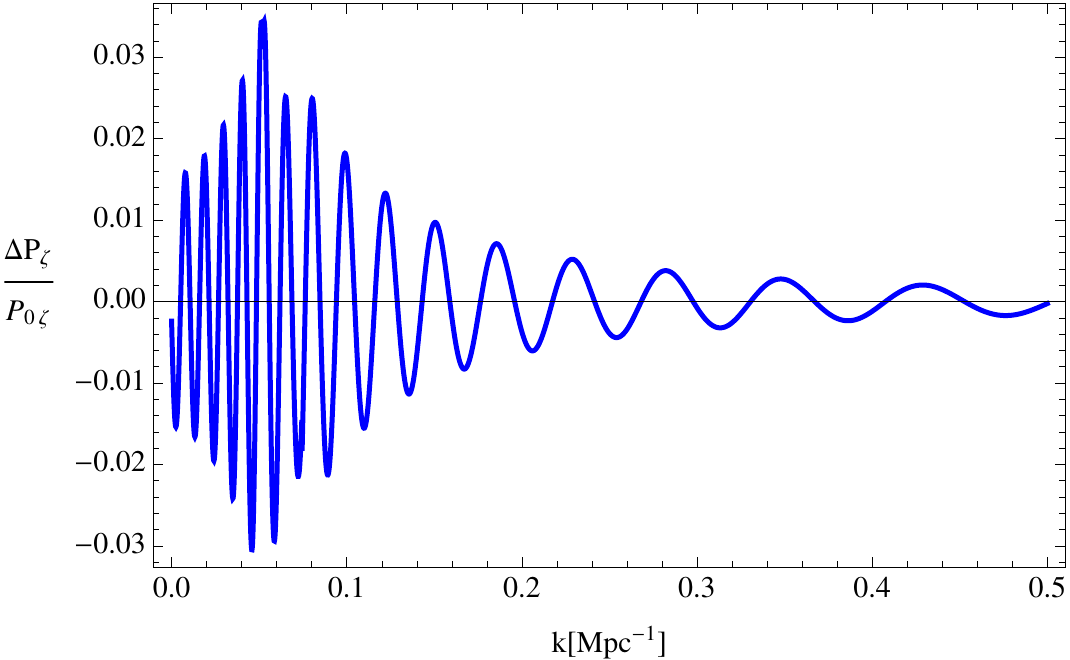}
\hspace{.5cm}
\caption{\small The full standard clock signal (Eq.\eqref{Template_SC_full}) with $k_r=0.1086/$Mpc and $C=0.0307$.}
\label{Fig:SC}
\end{figure}

Currently, a general analytical form for the full standard clock signal is unavailable. A special example of full clock signal related to the best-fit inflationary clock signal (see Sec.~\ref{Sec:clock}) is constructed in \cite{Chen:2014joa,Chen:2014cwa}. It was obtained by first finding the best-fit model for the inflation clock signal \eqref{Template:inf_clock}.
This determines all the parameters of the model and its full prediction can be simulated numerically.
The resulting full clock signal is then mimicked by the following template,
\bea
\frac{\Delta P_\zeta}{P_{\zeta0}} =
\left\{
\begin{array}{ll}
\high{
C \left[ 7\times 10^{-4} \left( \frac{2k}{k_0} \right)^2 + 0.5 \right]
\cos \left[ \frac{2k}{k_0} + 0.55\pi \right] ~,
}
&
k < k_a ~,
\\
\high{
\frac{14}{13} C \left( \frac{2k}{k_r} \right)^{-3/2}
\sin \left[ \Omega \ln (\frac{2k}{k_r}) + 0.75\pi \right] ~,
}
&
k_b >k \ge k_a ~,
\\
\high{
\frac{19}{13} C \left( \frac{2k}{k_r} \right)^{-3/2}
\sin \left[ \Omega \ln (\frac{2k}{k_r}) + 0.75\pi \right] ~,
}
&
k \ge k_b ~,
\end{array}
\right.
\label{Template_SC_full}
\eea
where
\bea
k_0 = \frac{k_r}{1.05\Omega} ~, ~~~
k_a = \frac{67}{140}k_r ~, ~~~
k_b = \frac{24}{35}k_r ~, ~~~
\Omega=30 ~,
\eea
which has only two free parameters and we choose $ k_r = 0.109/$Mpc and $C=0.0307$ (these are the best-fit values obtained in \cite{Chen:2014cwa} for a candidate with marginal significance).
The drawback of this procedure of finding full clock templates is that some parameters have to be fixed in the process. Ideally the full clock signal template should have the same number of parameters as the clock signal template.
In Fig. \ref{Fig:SC} we plot the template with these best-fit values.

\subsection{Bump feature}
\label{Sec:local}
Although the main goal of this paper is to study the oscillatory features, we also study a non-oscillatory feature for comparison. Such a feature is mostly a bump in the momentum space.

\begin{figure}
\center
\includegraphics[scale=.8]{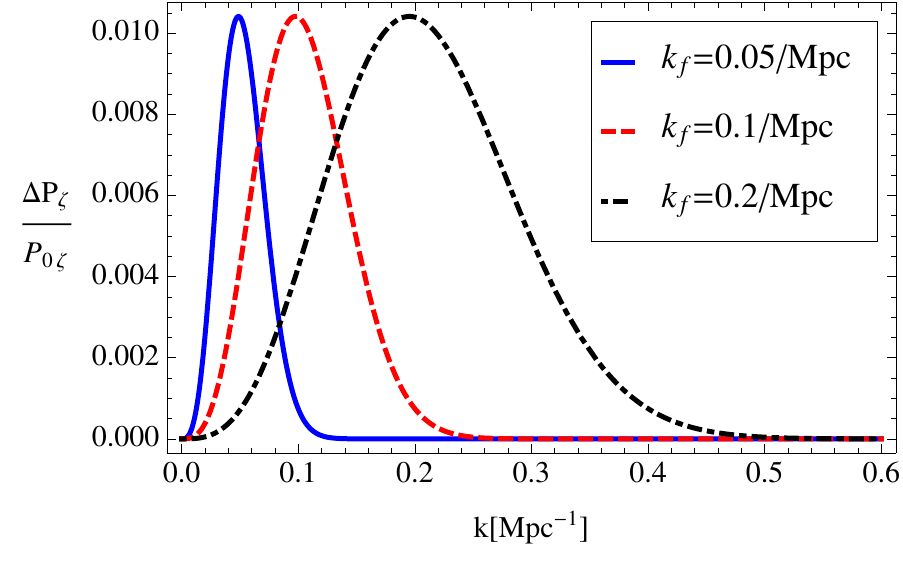}
\hspace{.5cm}
\caption{\small Bump feature at different locations in Fourier space with amplitude $C=0.01$.}
\label{Fig:local}
\end{figure}

There are many models in the literature that generate different bump-like features, each of which has its own motivation. Here we investigate the following template, which has been extensively studied as a feature generated by particle production mechanisms during inflation \cite{Barnaby:2009dd,Barnaby:2009mc,Barnaby:2010ke,Chantavat:2010vt,Romano:2008rr}:
\ba
\label{Template:local}
\dfrac{\Delta P_\zeta}{P_{\zeta_0}} = C \left( \dfrac{\pi e}{3} \right)^{3/2} \left( \frac{k}{k_f}\right)^3 e^{-\frac{\pi}{2}(\frac{k}{k_f})^2},
\ea
where $C$ is the amplitude of the feature and $k_f$ determines the location of the feature. In this work we choose  $C=0.01$ for the amplitude and we explore different values for $k_f$: $k_f=0.05{\rm Mpc}^{-1}$, $k_f = 0.1 {\rm Mpc}^{-1}$ and $k_f=0.2 {\rm Mpc}^{-1}$ for the position of the feature in k-space (see Fig. \ref{Fig:local}).

A summary of all models and templates that are studied in this work is in  Table   \ref{Table:models}.

Before closing this section we comment that almost all classes of  models discussed above have a potential to generate a sizable non-Gaussianity as well. Although the amplitude of the non-Gaussian signal is usually an independent parameter, the non-Gaussianities and the power spectrum signal have highly correlated oscillatory scale-dependence. In this case, a joint analysis of the two- and three-point function is required, which is an interesting topic of research. This type of analysis is beyond the scope of this paper. For the present application we always assume that the amplitude of the non-Gaussian signal is small enough so that it can be neglected in our investigation.

\begin{table}[t]
  \begin{center}
\begin{tabular}{|c|c|c|c|c|} \hline
Section & Model & Template & Parameters  and fiducial values  & Constraints\\
\hline \hline
\ref{Sec:Sharp} & Sharp feature & Eq.\eqref{Template_Sharp} & $C=0.03$, $\phi=0$ & Tab.~\ref{Table:sin}\\

 &  &  & $k_f [{\rm Mpc}^{-1}] = (0.004, 0.03, 0.1)$ & \\
\hline
\ref{Sec:step} & Step in $V(\phi)$ & Eq.\eqref{Template:step} & $\calC = 0.218$, $\eta_f = 1.44$Gpc, $x_d=1.6$ & Tab.~\ref{Table:step}\\
\hline
\ref{Sec:Resonance} & Resonance feature & Eq.\eqref{Template_Resonance} & $C=0.03$, $\Omega = (5, 30, 100)$, $\phi=0$ & Tab.~\ref{Table:res}\\
\hline
&Clock signal& Eq.\eqref{Template:clock} & $C=0.05$, $p=(2/3,1/5,-1/5)$,   & Tab.~\ref{Table:alt_clock}
\\ & for alternatives&& $k_r = 0.1$/Mpc, $\Omega=100$, $\phi=0$  & \\
 \ref{Sec:clock}&&&& \\
& Clock signal  &  Eq.\eqref{Template:inf_clock} & $C=0.0576$, $k_r = 0.108/$Mpc,  & Tab.~\ref{Table:inf_clock}
\\
&for inflation& & $\Omega=59.1$, $\phi=4.07$ & \\
\hline
\ref{Sec:SC_full} & Full standard clock & Eq.\eqref{Template_SC_full} & $C=0.0307$, $ k_r[{\rm Mpc}^{-1}] =0.109$ & Tab.~\ref{Table:SC} \\
\hline
\ref{Sec:local} & Bump feature & Eq.\eqref{Template:local} & $C=0.01$, $k_f [{\rm Mpc}^{-1}]=0.05, 0.1, 0.2$ & Tab.~\ref{Table:local}
\\
\hline
\end{tabular}
 \caption{\small A summary of all models studied in this work including  the free parameters for each model and  their templates, as well as a pointer to the table reporting their forecasted constraints. For the step model we choose the best fit values obtained in \cite{Adshead:2011jq}. The best fit values for the inflationary clock signal and standard clock signal are obtained in \cite{Chen:2014cwa}. In addition, one sharp feature model (with $k_f=0.004$) and one resonance feature model (with $\Omega=30$) are chosen to approximately represent the best-fit models found in \cite{Adshead:2011jq,Chen:2014cwa}.
 In the absence of the data analysis for other models, we choose fiducial values for parameters such that they are theoretically and observationally interesting, while roughly consistent with the null observation of signals in currently available data. }
\label{Table:models}
\end{center}
\end{table}

\section{Fisher matrix analysis}
\label{sec:Fisher}

To forecast expected errors around the fiducial model parameters presented above, we use the Fisher matrix approach, which is the standard workhorse for this.  The Fisher matrix is formed with the second derivatives with respect to the parameters  of the likelihood function around its maximum. The covariance matrix for the parameters is then  the inverse of the Fisher matrix, and parameter errors are estimated from the covariance matrix. Of course, if the likelihood of the model is not well approximated by a multi-variate Gaussian,   this procedure will yield a bad approximation of the error. Note that even if the likelihood for the data is a multi-variate Gaussian, the likelihood  as a function of the parameters may still be non-Gaussian if the model dependence on the parameters is non-linear.
For oscillatory features, where the likelihood may be multi-peaked and therefore highly non-Gaussian,  the result of a Fisher analysis  should be interpreted with care.  In fact, the Gaussian approximation of the likelihood around a (non unique) maximum, can yield artificially small error-bars and miss alternative ``solutions'' (``likelihood islands'') in parameter space. This is particularly important  in the low signal-to-noise regime. Here we still adopt the Fisher matrix approach because of its simplicity and speed. We note that  the results we report are relatively high signal-to-noise and their robustness  is checked by ensuring stability as function of the adopted step size in the computation of numerical derivatives. Of course, the present analysis applies only in the region of parameter space around the global maximum of the likelihood and  does not uncover or explore other local maxima which offer ``alternative solutions".  Nevertheless, the results presented here are an estimate of the  statistical power of the adopted survey to detect  features  and constrain their properties.


After computing the Fisher matrix ($F$ with components $F_{ij}$) the marginalized constraint on each parameter, say $x$, is  $\sqrt{(F^{-1})_{xx}}$.
 The joint marginalized constraint for a pair of parameters  can also be obtained by the following procedure \cite{Verde:2007wf,Verde:2009tu,Coe:2009xf}. Let us denote the two parameters $x$ and $y$ with fiducial values $\bar x$ and $\bar y$. Using the inverse of the  Fisher matrix we can construct the following $2 \times 2$ matrix:
\ba
Q^{-1} =
\left(
\begin{tabular}{cc}
$(F^{-1})_{xx}$ &$ (F^{-1})_{xy}$
\\
$(F^{-1})_{xy}$ & $(F^{-1})_{yy}   $
\end{tabular}
\right).
\ea
The inverse of the above matrix determines the  equation of the ellipse of the marginalized constraint, that is:
\ba
Q_{xx} \, (x-\bar x)^2+ 2 Q_{xy} \, (x-\bar x) \, (y-\bar y) + Q_{yy} \, (y-\bar y)^2 = \delta \chi^2,
\ea
in which $\delta \chi^2 = 2.3$ and $6.17$ correspond to the 1-$\sigma$ and 2-$\sigma$  confidence, respectively.

\subsection{Modeling the power spectrum signal and computing the Fisher matrix}

For the Fisher matrix analysis we use the publicly available software ``cosmology object oriented  package" COOP\footnote{Code home page with documentation at  http://www.cita.utoronto.ca/$\sim$zqhuang/coop/, direct download from https://github.com/zqhuang/COOP} \cite{Huang:2015srv}, which was also at the core of Ref. \cite{Huang:2012mr}. Here we briefly summarize  the essential information  relevant to the present application, and we refer the reader to the code documentation for details.

To compute the  (derivatives of the) likelihood function, we have to relate
 the primordial power spectrum to the observed galaxy power spectrum.
 The relation between the primordial power spectrum and the dark matter power spectrum, which can be readily derived from the Poisson and the Euler equations, is:
\ba
P(\bfk) =\dfrac{8 \pi^2}{25} k \, \dfrac{T(k)^2 \, D(z)^2}{\Omega_m^2 H_0^4} \, \calP_\zeta({\bf k}),
\label{Eq:Pk}
\ea
where $P(k)$ is the matter power spectrum, and $\Omega_m$ and $H_0$ are present time matter density fraction and Hubble parameter. $T(k)$ is the transfer function, normalized to one at large scales and $D(z)$ is the growth factor normalized to one at present time ($D(0)=1$). The transfer function and the growth factor can be computed numerically for a $\Lambda$CDM Universe, by using e.g., CAMB \cite{Lewis:1999bs}.

 The redshift-space galaxy power spectrum $P_g$ is then related to the (real space) matter power spectrum by \cite{Kaiser:1987, Peacock92}:
\ba
\label{Eq:Pg}
P_g(k,\mu) = \left[ 1+\beta \mu^2   \right]^2 b^2 P(k) \, e^{-c^2 \sigma_z^2 k^2 \mu^2 /H^2},
\ea
where $\mu$ denotes the cosine of the angle with respect to the line of sight, $b$ is the (linear) galaxy bias, $\beta=f/b$, and $f=d \ln \delta/d \ln a \sim \Omega_m^{0.56}$. This equation is strictly valid only in the limit of flat sky, or distant observer approximation,  but for our  forecasts it suffices.
In Eq. (\ref{Eq:Pg}) the first factor accounts for the linear redshift distortions (also known as {``Kaiser effect"}).
The exponential factor in the right hand side of Eq.\eqref{Eq:Pg} accounts for the smearing due to non-linear redshift distortions and uncertainty in redshift determination.
Here $\sigma_z$ includes the effects of both the intrinsic galaxy velocity dispersion and how well the redshifts are measured,
\ba
\sigma_z^2 = (1+z)^2 \left[ \sigma_{0\gamma z}^2 + \sigma_{0v}^2  \right]~,
\ea
where $\sigma_{0\gamma z}$ and $ \sigma_{0v}$ correspond to photometric/spectroscopic-redshift error and virialized motion of galaxies, respectively, as if the galaxies were located at redshift zero.
The virialized {\it r.m.s.} velocity of galaxies is approximated by the conservative value $\sigma_v/(1+z) \simeq 560 \pm 280 \, $ km/s (see e.g., \cite{Huang:2012mr,Huang:2015srv} and COOP code documentation), which corresponds to $\sigma_{0v} =  0.0019 \pm 0.0009$ and we marginalize over the reported uncertainty assuming a Gaussian distribution.
The photometric/spectroscopic-redshift error, $\sigma_{0\gamma z}$, depends on the details of the observation (as we discuss below).  Unlike the $\sigma_{0v}$, we do not marginalize over $\sigma_{0\gamma}$ because the latter error is in principle instrumental and can be estimated with a good accuracy. We will see that for a photometric observation (like LSST) the main error comes from redshift uncertainty, i.e., $\sigma_{0\gamma z} \gg  \sigma_{0v}$; while for the spectroscopic observation (like Euclid) these two effects are comparable, i.e. $\sigma_{0\gamma z} \sim  \sigma_{0v}$.
 In  reality the redshift dependence of uncertainty in the  redshift determination  could well be different from that of the  intrinsic galaxy velocity dispersion.  For these forecasts as a first approximation to estimate their impact on the final error-bars,  these  two effect are assumed to  be fully degenerate. In practice the two effects may even be separable if they have a different $z$ dependence.

There is yet another real-world effect which we need to take into account. The volume of the survey is finite, hence the k-resolution and sampling  is not infinite. This effect can be simulated by smoothing the power spectrum via a Gaussian window function as follows \cite{Huang:2012mr},
\ba
\hat{P}({\bf k}) = \int d^3 {\bf k} \, P({\bf k'}) \vert  W(\vert \bfk-\bfk' \vert)\vert^2,
\label{Eq:smooth}
\ea
in which
\ba
\vert W(k) \vert^2 = \dfrac{1}{(2\pi \sigma_W^2)^{3/2}} \exp\left(-\dfrac{k^2}{2\sigma_W^2}\right),
\ea
where $\sigma_W$ is the width of the window function, for which we choose:
\ba
\sigma_W = \dfrac{\sqrt{2\ln 2}}{2 \pi} k_{\rm min},
\ea
where $k_{\rm min} = 2\pi  (3 V/4\pi)^{-1/3}$ is the minimum wave number which can be probed by the survey of volume $V$\footnote{Our choice of the minimum $k$ may be somewhat optimistic. While the size of the survey fundamentally limits the largest scale that can be sampled, systematic  and calibration effects may in  practice  reduce this further, increasing the $k_{min}$ that can reliably be measured. These effects are however very hard to anticipate  and doing so goes well beyond the scope of this paper. It would suffice here to note that because of the large cosmic variance on large scales our constraints degrade ``gracefully" with increasing in $k_{min}$ as long as the survey volume is not affected significantly.
}.
This choice of the width for the window function makes the half-height real space window function to correspond to the total volume \cite{Huang:2012mr}. The above smoothed power spectrum has to be computed at each redshift, or roughly, for each redshift bin as long as the bins are sufficiently narrow.

As a result,  the high frequency oscillatory features are smeared out making it more difficult to discriminate among different models with slightly different frequencies.
  For the sinusoidal feature it is easy to check that as long as the width of the window function $\sigma_W$ is smaller than the frequency of the oscillations, the effect of the window function can be neglected. This is indeed the case here for the sharp feature signal with the specific fiducial values that we are going to investigate (see Table \ref{Table:models} and compare with Table \ref{Table:bins_LSST} and \ref{Table:bins_Euclid}). However, we note that for a more complicated template this approximation does not hold and one can expect a significant effect, depending on the model and the fiducial values.

 The Fisher matrix for a galaxy survey can be written as follows  e.g., \cite{Tegmark:1997rp}:
\ba
F_{ij}& =& \dfrac12 \int_{\bfk_{\rm min}}^{\bfk_{\rm max}}
\dfrac{\partial \ln \hat P_g(\bfk)}{\partial \theta_i} \dfrac{\partial \ln\hat  P_g(\bfk)}{\partial \theta_j} V_{\text{eff}}(\bfk)\dfrac{d^3 k}{(2\pi)^3}
\\
&=&\int_{-1}^{1} \int_{k_{\rm min}}^{k_{\rm max}} \dfrac{\partial \ln \hat P_g(k,\mu)}{\partial \theta_i} \dfrac{\partial \ln \hat P_g(k,\mu)}{\partial \theta_j} V_{\rm{eff}}(k,\mu)\dfrac{k^2dk d\mu}{2(2\pi)^2},
\label{Eq:fisher}
\ea
where $\theta_i$ are the free parameters of the model. We set $k_{\rm min}$ to be the largest scale that can be probed by the survey, and $k_{\rm max}$ to the smallest scale below which the linear power spectrum is contaminated by non-linearities (see below for a more rigorous definition).
In the above equation, $V_{\text{eff}}$ is the effective volume of the survey defined by:
\ba
V_{\text{eff}} = \int \left[ \dfrac{n({\bf{r}})  \hat P_{g}(k,\mu)}{n({\bf{r}}) \hat P_{g}(k,\mu)+1} \right]^2
d^3 r
\simeq
\left[ \dfrac{\bar n  \hat P_{g}(k,\mu)}{\bar n \hat P_{g}(k,\mu)+1} \right]^2V,
\label{Eq:Veff}
\ea
where $\bar n$ the mean number density of galaxies and in the second equality we neglected the position dependence of the number density $n({\bf r})$.

 For a survey covering a fraction of the sky $f_{\rm sky}$ the volume is:
\ba
V=  \dfrac{4 \pi}{3} \times f_{\rm sky}
  \left[d_c(z_{\rm max})^3-d_c(z_{\rm min})^3\right],
  \label{Eq:V}
\ea
and $d_c(z)$ is the comoving distance from redshift $z$:
\ba
d_c(z) = \int_0^z \dfrac{c}{H(z)} dz.
\ea

For a single redshift bin, the volume in
Eq. \eqref{Eq:Veff}, $k_{\rm min}$ and $k_{\rm max}$ in Eq. \eqref{Eq:fisher} have to be replaced by the corresponding parameters for the bin. Then the full Fisher matrix is the sum over all Fisher matrices for each bin (i.e., we assume uncorrelated redshift bins). Once the volume of the survey for each bin is determined by Eq.\eqref{Eq:V}, then each $k_{\rm min}$ can be computed as $k_{\rm min} =  2\pi (3V/4\pi)^{-3}$.

For setting  the $k_{\rm max}$ values,  we first estimate the {\it r.m.s} value of fluctuations at  scale $R$ and redshift $z$, by smoothing the power spectrum via a top-hat window function corresponding to a sphere with radius $R$ in real space:
\ba
\sigma^2_R(z) = \int \dfrac{d^3k}{(2\pi)^3} P(k,z) W_{\rm th}^2(k R),
\ea
where $P(k,z)$ is the matter power spectrum (Eq.~\eqref{Eq:Pk}), and $W_{\rm th}(kR)$ is the Fourier transform of the top-hat window function,
\ba
W_{\rm th}(x) = 3 \dfrac{\sin(x)-x \cos(x)}{x^3}.
\ea
The wavenumber $k_{\rm max}$ is then determined by identifying $k_{\rm max}=\pi/2/R_*$ and requiring $\sigma_{R_*}(z) = 1/2$. We choose the threshold of {\it r.m.s.} density fluctuation to be $1/2$ to guarantee that all the scales considered are within the linear regime. The signal-to-noise can be further improved if non-linearities are accurately modeled; in this sense our forecast is mildly conservative.

\subsection{Adopted survey specifications}


Our aim is to forecast the constraints that a photometry (LSST-like) as well as a spectroscopy (Euclid-like) observation would be able to place on our parameters. In this subsection we quantify how we would mimic the real observations.

The LSST experiment  \cite{LSST}  will observe $23,000$ square degrees of the sky ($f_{\rm sky}=0.58$) in the redshift range $z=[0.2-3]$. The photometric redshift error will be $\sigma_{\gamma z} = \sigma_{0\gamma z} (1+z)$ with $\sigma_{0\gamma z} = 0.04$. The bias parameter is  modeled by $b=1+0.84 z$ \cite{Zhan:2005ki}.  The {\it{surface}} number density depends on the redshift; we will consider here the following relation \cite{Zhan:2005ki}:
\ba
n_s(z) = 640 \, z^2 \, e^{-z/0.35} \, \, \text{arcmin}^{-2},
\ea
which corresponds to a total projected number density of:
\ba
\int_{0}^{\infty} n_s(z) dz \simeq 55 \, \text{arcmin}^{-2}.
\ea
The total number density for the survey (and similarly for each redshift bin) can be computed as:
\ba
\bar n =\dfrac{4\pi f_{\rm sky}}{V}\int_{z_{\rm min}}^{z_{\rm max}} n_s(z) dz  .
\ea

 Following Ref. \cite{Zhan:2005ki}, we divide the survey volume into seven redshift bins with roughly the same comoving radial intervals. The mean redshifts, the redshift range and the $k_{\rm min}$ and $k_{\rm max}$ for each bin are summarized in Table \ref{Table:bins_LSST}.

\begin{table}[t]
  \begin{center}
\begin{tabular}{|c|c|c|c|c|} \hline
$z_{\rm mean}$ & $z$-range & $\bar n [h^{3} {\rm Mpc}^{-3}]$ &$k_{\rm min}[h/$Mpc] & $k_{\rm max}[h/$Mpc]\\
\hline
0.31 & 0.2 -- 0.46 & 0.15& 0.0071 & 0.08
\\
0.55 & 0.46 -- 0.64 &0.10 & 0.0050 & 0.09
\\
0.84 & 0.64 -- 1.04& 0.064 &0.0040 & 0.11
\\
1.18 & 1.04 -- 1.32 &0.036 &0.0035 & 0.14
\\
1.59 & 1.32 -- 1.86&0.017 &0.0030 & 0.17
\\
2.08 & 1.86 -- 2.3 & 0.0069&0.0028& 0.23
\\
2.67 & 2.3 -- 3 & 0.0022 &0.0026 & 0.31
\\
\hline
\end{tabular}
 \caption{\small The list of seven redshift bins and $k_{\rm min}$ and $k_{\rm max}$ for each bin for a LSST-like observation. The surface number density is assumed to be $n_s(z) = 640 \, z^2 \, e^{-z/0.35} \, \, \text{arcmin}^{-2}$, the bias $b=1+0.84 z$, sky coverage $f_{\rm sky}=0.58$, and photometric-redshift error $\sigma_{0\gamma z}=0.04$.}
\label{Table:bins_LSST}
\end{center}
\end{table}

\begin{table}[t]
  \begin{center}
\begin{tabular}{|c|c|c|c|c|} \hline
$z_{\rm mean}$ & $z$-range & $n_{\rm obs} [h^{3} {\rm Mpc}^{-3}]$ & $k_{\rm min}[h/$Mpc] & $k_{\rm max}[h/$Mpc]\\
\hline
0.6 & 0.5 -- 0.7 & $3.56 \times 10^{-3}$& 0.0061 & 0.09
\\
0.8 & 0.7 -- 0.9 & $2.82 \times 10^{-3}$& 0.0054 & 0.11
\\
1.0 & 0.9 -- 1.1& $1.81 \times 10^{-3}$& 0.0051 & 0.12
\\
1.2 & 1.1 -- 1.3 & $1.44 \times 10^{-3}$& 0.0048 & 0.14
\\
1.4 & 1.3 -- 1.5& $0.99 \times 10^{-3}$& 0.0047 & 0.16
\\
1.6 & 1.5 -- 1.7 & $0.55 \times 10^{-3}$& 0.0046& 0.18
\\
1.8 & 1.7 -- 1.9 &$ 0.29 \times 10^{-3}$&0.0045 & 0.20
\\
2.0 & 1.9 -- 2.1& $0.15 \times 10^{-3}$ & 0.0045 &0.22
\\
\hline
\end{tabular}
 \caption{\small The list of eight redshift bins, number densities of to-be-observed galaxies $n_{\rm obs}$ and $k_{\rm min}$ and $k_{\rm max}$ for each bin for a Euclid-like observation. The number density used in our analysis is: $\bar n=\epsilon \, n_{\rm obs}$, for which we choose $\epsilon=0.5$. The bias is $b=\sqrt{1+z}$, sky coverage $f_{\rm sky}=0.36$, and spectroscopy-redshift error $\sigma_{0\gamma z}=0.001$.}
\label{Table:bins_Euclid}
\end{center}
\end{table}

As for the Euclid survey, we take the information about the redshift bins and the number densities from \cite{Amiaux:2012bt, Laureijs:2011gra}; see Table \ref{Table:bins_Euclid} for a summary. As for the number densities, following ref. \cite{Amendola:2012ys}, we take the efficiency $\epsilon=0.5$, hence we use the number density $\bar n=\epsilon \, n_{\rm obs}$ in our analysis. For the bias parameter we choose $b=\sqrt{1+z}$. Note that the difference between the bias parameters in Euclid and LSST as a function of redshift is due to the different types of galaxies that each survey is going to target.

In our analysis, a CMB prior is imposed  by adding to the  Fisher matrix for LSS a Fisher matrix for  the Planck mission. The implementation follows Sec 2A of Ref. \cite{Huang:2012mr}.
For estimating the noise  we use Planck's 70 GHz, 100 GHz and 143 GHz channels, while we assume that other channels are used to remove the foreground effect. We also take into account the uncertainty of noise power spectrum estimation, for which we assume a $1\%$ accuracy. Furthermore, we do not consider the  polarization signal at high $\ell$. Our forecasted ``Planck" only error-bars match those  obtained and published by the Planck collaboration in the 2015 data release, for both standard cosmological parameters and for feature models.\footnote{The Planck data release of 2015 include  high $\ell$ polarization data, but it is not clear how much information on features polarization adds because the polarization data is used in band powers   and the  process of  foreground subtraction  and the treatment of  other real world effect effectively increases the noise.}

In what follows, we adopt  the following fiducial values for the parameters in our Fisher forecast: ($A_s$, $n_s$, $\Omega_c h^2$, $\Omega_b h^2$, $h$, $\tau$ ) $=( 2.15 \times 10^{-9},  0.969, 0.12, 0.022, 0.67, 0.07)$, which are consistent with Planck data (see Table \ref{Table:models} for the fiducial values of the feature parameters). Note that, besides all the cosmological and primordial parameters, we allow the bias parameter for each bin to vary, and then marginalize over them to obtain constraints on primordial parameters.

\section{Results and discussions}
\label{sec:results}

Before we present the results from the Fisher forecast analysis let us make several general comments:
\begin{itemize}
\item In  the tables, ``Planck" approximates the current constraints from the Planck 2015 data, in which we used the TT and low-$\ell$ polarization. Polarization information  has not yet  played an important role in these constraints. ``+LSST" and ``+Euclid" forecasts, respectively, refer to the constraints that can be obtained by adding the data from the LSST-like and Euclid-like experiments (specified in Sec.~\ref{sec:Fisher}), respectively, to the Planck data.

\item Among various parameters, the amplitude of the feature $C$ is the most important. Once the amplitude is detected with high statistical significance, all other parameters are also constrained  in a statistically significant way. Note that it is possible that, for some fiducial values, other parameters such as the frequency can appear to be statistically significant even if $C$ is not (e.g.~the 3rd example in Table~\ref{Table:res}); one cannot claim detection in such cases.

\item Although in this work we assume non-zero fiducial values for the feature signals (in some cases, the amplitude is even too large to be compatible with the current constraints), as long as $C$ is small (as it is indeed  the case for the values considered here)  the features remain  small perturbations of the power spectrum, and the errors on $C$ remain approximately invariant to changes of the fiducial values. Note that  this is not necessarily the case for the other parameters, of which the errors  may change for different choices of  fiducial values. Therefore, it is worth to emphasize that the most important results of the paper, namely the projected errors for the amplitudes $C$, also apply to other fiducial amplitudes. (See also the comments at the beginning of Sec.~\ref{sec:Fisher}.)

\item For convenience, the template, fiducial parameters, and table of results associated with each class of models are listed in Table.~\ref{Table:models}.

\end{itemize}

\begin{figure}[h]
\center
\includegraphics[scale=.7]{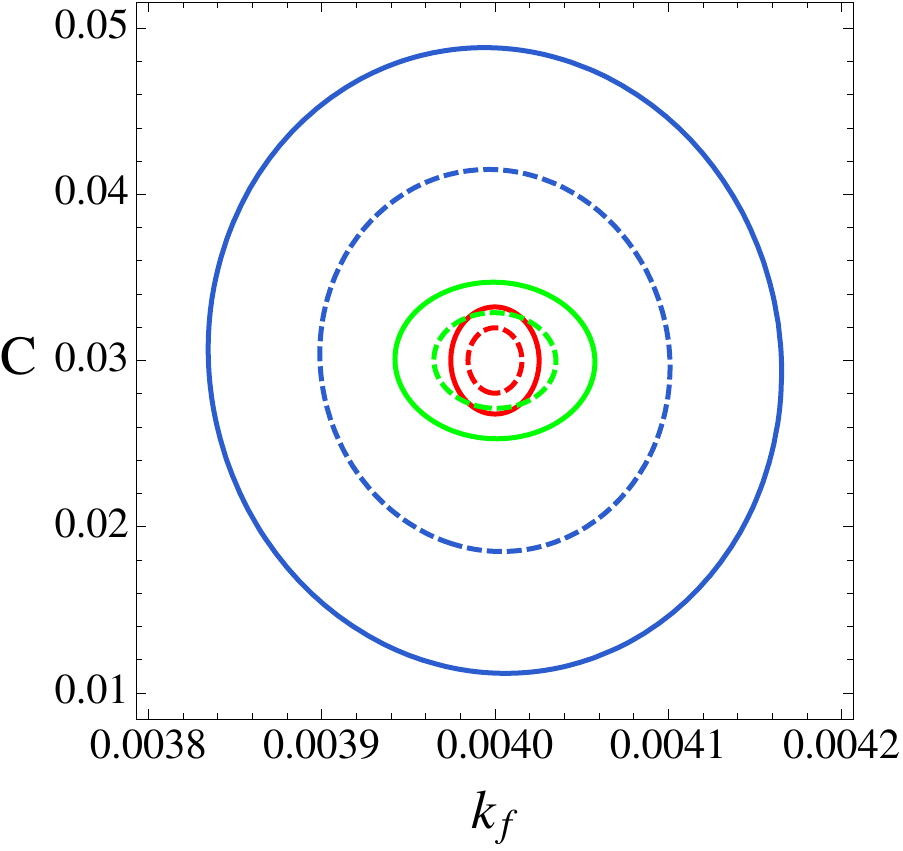}
\caption{\small ({\bf Sharp feature}) Joint   2D constraint ($k_f$ vs. $C$) for a sharp-feature signal with $k_f=0.004$/Mpc, $C=0.03$ and $\phi=0$ from Planck (blue), Planck+ LSST-like (red), and Planck+ Euclid-like (green) observations. Solid (dashed) lines enclose the $2-\sigma$ ($1-\sigma$) confidence regions.}
\label{Fig:2d_sin}
\end{figure}

In Table \ref{Table:sin} we present the marginalized $1-\sigma$ constraints for the sharp feature template, each sub-table correspond to a different fiducial value of the parameter $k_f$.  The joint $k_f,\, C$,  $1-\sigma$ and $2-\sigma$ contours for a case with $k_f=0.004$ ${\rm Mpc}^{-1}$ are shown  in Fig.~\ref{Fig:2d_sin}.
Hereafter in the figures, the dashed contours correspond to $1-\sigma$ confidence and solid contours to $2-\sigma$; blue denotes ``Planck", green  when adding ``Euclid" and red  when combining ``Planck" with ``LSST". Constraints are always marginalized over all remaining parameters.

In Table \ref{Table:step} we report the  marginalized  $1-\sigma$ constraints for a model with a step in the potential. As mentioned in Sec.~\ref{Sec:step}, this is a special example of the sharp feature class for which the envelop behavior is important and the template we use is not a good approximation.

\begin{figure}
\center
\includegraphics[scale=.5]{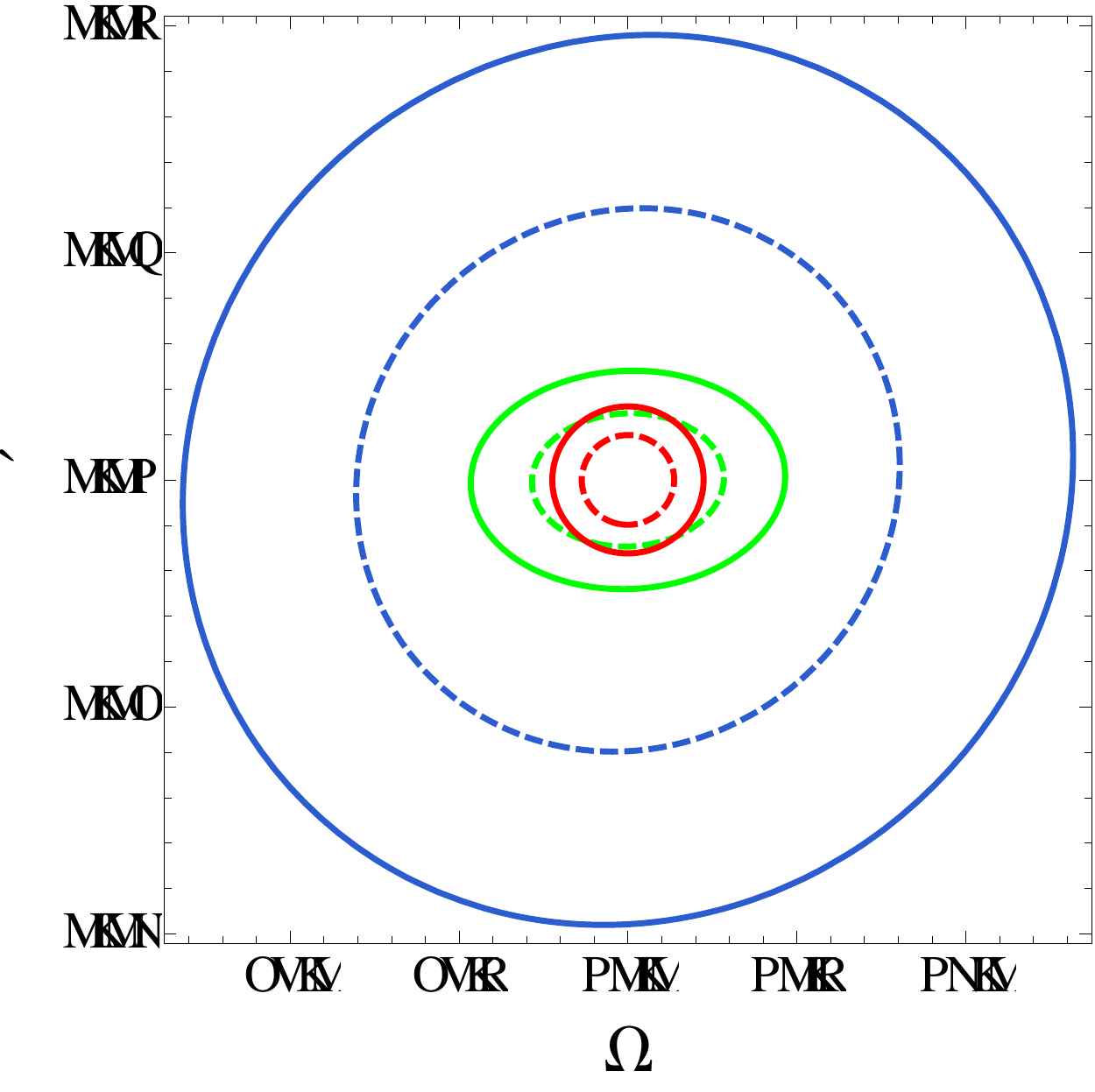}
\caption{\small ({\bf Resonance feature}) Joint $\Omega,\,C$ constraints for a resonance signal  with $C=0.03$, $\Omega=30$ and $\phi=0$. The blue, the red, and the green contours are respectively for Planck, Planck+LSST-like and Planck+Euclid-like observations.  Solid (dashed) lines enclose the $2-\sigma$ ($1-\sigma$) confidence regions.}
\label{Fig:2d_res}
\end{figure}

In Table \ref{Table:res} we present the marginalized  $1-\sigma$ constraints for the resonance feature template;  the  $(\Omega,\,C)$ \, 1 and 2$-\sigma$ joint   contours for a case with $\Omega=30$  are shown in Fig.~\ref{Fig:2d_res}.

\begin{table}[t]
  \begin{center}
\begin{tabular}{|c||c|c|c|c|c|}
\multicolumn{1}{l}{}\\
\multicolumn{4}{r}{$k_f =0.004/{\rm Mpc} \hspace{-.3cm}$}
\\
\cline{2-6}
\multicolumn{1}{c|}{}
&$A_s$ & $n_s$ & $C$ &$ k_f [{\rm Mpc}^{-1}]$ & $\phi$  \\
\hline
Planck & $7.3 \times 10^{-11}$ & 0.0056 & 0.0075 & 0.000065 & 0.51 \\ \hline\hline
+ LSST & $6.1 \times 10^{-11}$ & 0.0019 & 0.0013 & 0.000010 & 0.10 \\ \hline
+ Euclid & $3.2 \times 10^{-11}$ & 0.0031 & 0.0019 & 0.000023 & 0.19 \\ \hline
\multicolumn{1}{l}{}\\
\multicolumn{4}{r}{$k_f =0.03/{\rm Mpc} \hspace{-.3cm}$}
\\
\cline{2-6}
\multicolumn{1}{c|}{}
&$A_s$ & $n_s$ & $C$ &$ k_f[{\rm Mpc}^{-1}]$ & $\phi$  \\
\hline
 Planck & $7.6 \times 10^{-11}$ & 0.0060 & 0.0031 & 0.0021 & 0.38 \\ \hline \hline
 + LSST & $6.3 \times 10^{-11}$ & 0.0020 & 0.0012 & 0.00047 & 0.11 \\ \hline
 + Euclid & $3.2 \times 10^{-11}$ & 0.0036 & 0.0019 & 0.0013 & 0.21 \\ \hline
\multicolumn{1}{l}{}\\
\multicolumn{4}{r}{$k_f =0.1/{\rm Mpc} \hspace{-.3cm}$}
\\
\cline{2-6}
\multicolumn{1}{c|}{}
&$A_s$ & $n_s$ & $C$ &$ k_f[{\rm Mpc}^{-1}]$ & $\phi$  \\
\hline
 Planck & $2.17 \times 10^{-10}$ & 0.013 & 0.095 & 0.19 & 3.1 \\ \hline \hline
+ LSST & $6.8 \times 10^{-11}$ & 0.0070 & 0.012 & 0.031 & 0.65 \\ \hline
+ Euclid & $1.48 \times 10^{-10}$ & 0.0097 & 0.067 & 0.13 & 2.1 \\ \hline
\end{tabular}
  \caption{\small ({\bf Sharp feature}) Marginalized $1-\sigma$ constraints on the free parameters for sharp-feature models. The fiducial values for $C$ and $\phi$ are $0.03$ and  $0$, respectively, while $k_f$ is different for each sub-table. Each LSS survey is considered always in combination with ``Planck". See also Fig.\ref{Fig:2d_sin} for a joint 2D plot.}
\label{Table:sin}
  \end{center}
\end{table}

\begin{table}[t]
  \begin{center}
\begin{tabular}{|c||c|c|c|c|c|}
\cline{2-6}
\multicolumn{1}{c|}{}& $A_s$ & $n_s$ & $\calC$ &$\eta_f[{\rm Mpc}]$ &$x_d$  \\
\hline
 Planck & $7.3 \times 10^{-11}$ & 0.0057 & 0.22 & 230 & 1.8 \\ \hline \hline
+ LSST & $6.1 \times 10^{-11}$ & 0.0019 & 0.17 & 110 & 0.82 \\ \hline
+ Euclid & $3.2 \times 10^{-11}$ & 0.0032 & 0.21 & 230 & 1.7 \\ \hline
\end{tabular}
\caption{\small ({\bf An example of step-in-potential}) Marginalized $1-\sigma$ constraints on the parameters of the step in the potential model with the fiducial values $\calC = 0.218$, $\eta_f = 1.44$ Gpc and $x_d=1.6$. Each LSS survey is considered  in combination with ``Planck".}
\label{Table:step}
  \end{center}
\end{table}

\begin{table}[t]
  \begin{center}
\begin{tabular}{|c||c|c|c|c|c|}
\multicolumn{4}{r}{$\Omega =5 \hspace{.5cm}$}
\\
\cline{2-6}
\multicolumn{1}{c|}{}& $A_s$ & $n_s$ & $C$ &$ \Omega$ & $\phi$  \\
\hline
Planck & $7.5 \times 10^{-11}$ & 0.0068 & 0.0031 & 0.26 & 0.52 \\ \hline \hline
+ LSST & $6.2 \times 10^{-11}$ & 0.0021 & 0.0013 & 0.081 & 0.16 \\ \hline
+ Euclid & $3.2 \times 10^{-11}$ & 0.0035 & 0.0017 & 0.14 & 0.31 \\ \hline
\multicolumn{1}{l}{}\\
\multicolumn{4}{r}{$\Omega =30 \hspace{.4cm}$}
\\
\cline{2-6}
\multicolumn{1}{c|}{}
& $A_s$ & $n_s$ & $C$ &$ \Omega$ & $\phi$  \\
\hline
Planck & $7.3 \times 10^{-11}$ & 0.0064 & 0.0079 & 0.53 & 1.2 \\ \hline \hline
+ LSST & $6.1 \times 10^{-11}$ & 0.0019 & 0.0013 & 0.090 & 0.17 \\ \hline
+ Euclid & $3.2 \times 10^{-11}$ & 0.0031 & 0.0019 & 0.19 & 0.38 \\ \hline
\multicolumn{1}{l}{}\\
\multicolumn{4}{r}{$\Omega =100 \hspace{.2cm}$}
\\
\cline{2-6}
\multicolumn{1}{c|}{}
& $A_s$ & $n_s$ & $C$ &$ \Omega$ & $\phi$  \\
\hline
 Planck & $7.2 \times 10^{-11}$ & 0.0056 & 0.015 & 0.97 & 2.1 \\ \hline \hline
+ LSST & $6.1 \times 10^{-11}$ & 0.0019 & 0.0016 & 0.15 & 0.29 \\ \hline
+ Euclid & $3.2 \times 10^{-11}$ & 0.0031 & 0.0037 & 0.49 & 0.94 \\ \hline
\end{tabular}
 \caption{\small ({\bf Resonance feature}) Marginalized $1-\sigma$ constraints on the parameters of the resonance signal. The fiducial value for $\Omega$ is presented on top of each sub-table while we always have set $C=0.03$ and $\phi=0$. Each LSS survey is considered  in combination with "Planck". See Fig.\ref{Fig:2d_res} for a joint constraint on $C$ and $\Omega$.}
\label{Table:res}
  \end{center}
\end{table}

\begin{table}[t]
  \begin{center}
   \begin{tabular}{|c||c|c|c|c|c|c|c|}
\multicolumn{5}{r}{$p=2/3 \hspace{.3cm}$}
\\
\cline{2-8}
\multicolumn{1}{c|}{}& $A_s$ & $n_s$ & $C$ &$ k_r[{\rm Mpc}^{-1}]$ & $p$ & $ \Omega_{\rm eff}$ & $\phi$   \\
\hline
 Planck & $7.3 \times 10^{-11}$ & 0.0058 & 0.0062 & 0.0073 & 0.025 & 11.0 & 0.21 \\ \hline \hline
 +LSST & $6.1 \times 10^{-11}$ & 0.0019 & 0.0019 & 0.00077 & 0.0078 & 1.6 & 0.12 \\ \hline
 + Euclid & $3.2 \times 10^{-11}$ & 0.0032 & 0.0028 & 0.0010 & 0.011 & 1.9 & 0.16 \\ \hline
\multicolumn{1}{l}{}\\
\multicolumn{5}{r}{$p=1/5 \hspace{.3cm}$}
\\
\cline{2-8}
\multicolumn{1}{c|}{}& $A_s$ & $n_s$ & $C$ &$ k_r[{\rm Mpc}^{-1}]$ & $p$ & $ \Omega_{\rm eff}$ & $\phi$   \\
\hline
 Planck & $7.7 \times 10^{-11}$ & 0.0069 & 0.036 & 0.016 & 0.14 & 82 & 1.3 \\ \hline \hline
+ LSST & $6.2 \times 10^{-11}$ & 0.0023 & 0.0057 & 0.0018 & 0.020 & 8.2 & 0.32 \\ \hline
 + Euclid & $3.2 \times 10^{-11}$ & 0.0041 & 0.0074 & 0.0028 & 0.031 & 11 & 0.46 \\ \hline
\multicolumn{1}{l}{}\\
\multicolumn{5}{r}{$p=-1/5 \hspace{.3cm}$}
\\
\cline{2-8}
\multicolumn{1}{c|}{}& $A_s$ & $n_s$ & $C$ &$ k_r[{\rm Mpc}^{-1}]$ & $p$ & $ \Omega_{\rm eff}$ & $\phi$   \\
\hline
 Planck & $7.4 \times 10^{-11}$ & 0.0072 & 0.032 & 0.018 & 0.25 & 89. & 3.9 \\ \hline \hline
+ LSST & $6.2 \times 10^{-11}$ & 0.0021 & 0.0071 & 0.0019 & 0.040 & 12. & 0.85 \\ \hline
+ Euclid & $3.2 \times 10^{-11}$ & 0.0037 & 0.0085 & 0.0029 & 0.054 & 15. & 1.1 \\ \hline
\multicolumn{1}{l}{}\\
\end{tabular}
\caption{\small ({\bf Clock signals for alternative-to-inflation scenarios}) Marginalized 1-$\sigma$ constraints on the parameters of the clock signal for alternative scenarios. The fiducial values for $p$ are different for each sub-table, while for other parameters we have set $C=0.05$, $k_r =0.1$ Mpc$^{-1}$, $\Omega_{\rm eff}=100$, and $\phi=0$.  Each LSS survey is considered  in combination with ``Planck". See Fig. \ref{Fig:2d_clock_alt} for joint constraints.}
\label{Table:alt_clock}
  \end{center}
\end{table}

\begin{table}[t]
  \begin{center}
\begin{tabular}{|c||c|c|c|c|c|c|}
\cline{2-7}
\multicolumn{1}{c|}{}
&$A_s$ & $n_s$ & $C$ &$ k_r[{\rm Mpc}^{-1}]$ & $\Omega_{\rm eff}$ & $\phi$  \\
\hline
 Planck & $7.4 \times 10^{-11}$ & 0.0077 & 0.014 & 0.0042 & 3.3 & 3.4 \\ \hline \hline
 + LSST & $6.1 \times 10^{-11}$ & 0.0019 & 0.0030 & 0.00066 & 0.39 & 0.37 \\ \hline
 + Euclid & $3.2 \times 10^{-11}$ & 0.0032 & 0.0035 & 0.00085 & 0.72 & 0.70 \\ \hline
\end{tabular}
 \caption{\small ({\bf Clock signal for inflation}) Marginalized 1-$\sigma$ constraints on the free parameters of the inflationary clock signal. The fiducial values used for this table are $C=0.0576, k_r =0.108$Mpc$^{-1}$, $\Omega_{\rm eff}=59.1$ and $\phi = 4.07$.  Each LSS survey is considered  in combination with ``Planck". See Fig. \ref{Fig:2d_inf_clock} for a related contour plot.}
\label{Table:inf_clock}
\end{center}
\end{table}

\begin{figure}
\center
\includegraphics[scale=.65]{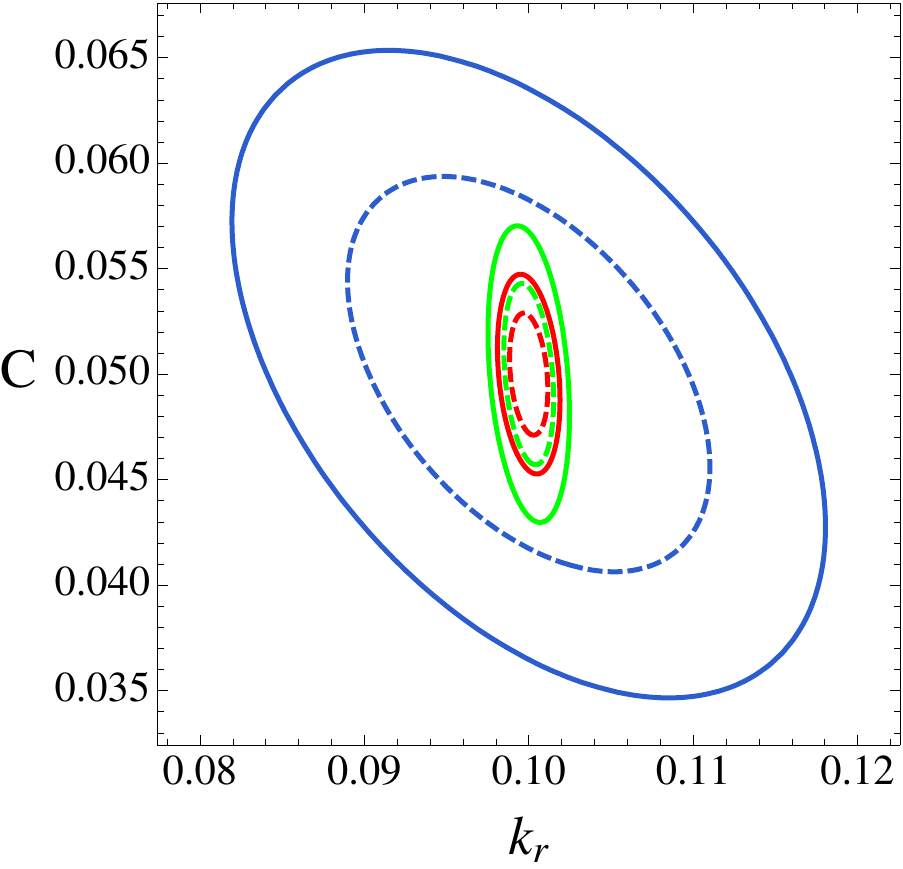}
\includegraphics[scale=.64]{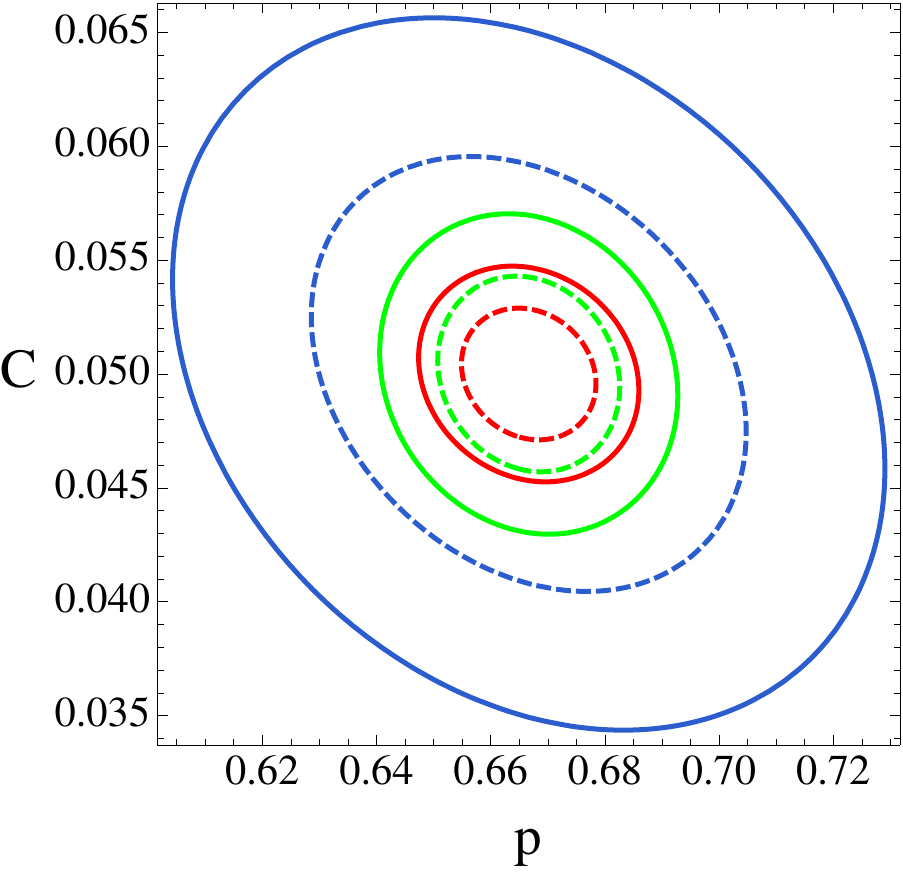}
\caption{\small ({\bf Clock signals for alternative-to-inflation}) Joint 1 and 2-$\sigma$ contours for the clock signal for the  matter contraction (fiducial value: $p=2/3$) scenario, Left: $k_r$ vs. $C$, right:  $p$ vs. $C$. Planck only constraints (blue),  Planck+LSST-like (red) and Planck+Euclid-like (green) experiments. Other parameters for the plots are the following: $(C, k_r, \Omega_{\rm eff}, \phi) = (0.05, 0.1/{\rm Mpc}, 100, 0)$. }
\label{Fig:2d_clock_alt}
\end{figure}

\begin{figure}
\center
\includegraphics[scale=.7]{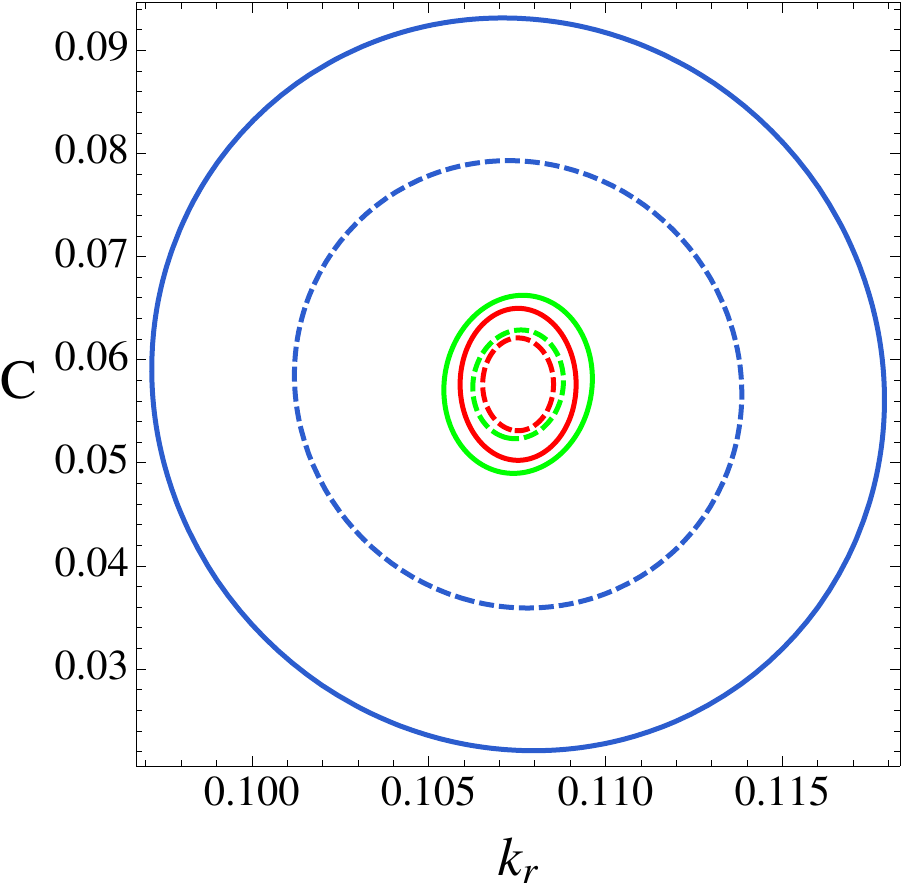}
\caption{\small ({\bf Clock signal for inflation})  Joint 1 and 2-$\sigma$ contours for the inflationary clock signal. The blue, red, and green contours are for Planck, Planck+LSST-like and Planck+ Euclid-like observations, respectively. The fiducial values are $C=0.0576$, $k_r=0.108$ Mpc$^{-1}$, $\Omega_{\rm eff}=59.1$ and $\phi=4.07$. }
\label{Fig:2d_inf_clock}
\end{figure}

\begin{table}[t]
  \begin{center}
\begin{tabular}{|c||c|c|c|c|}
\cline{2-5}
\multicolumn{1}{c|}{}&
$A_s$ & $n_s$ & $C$ &$k_r[{\rm Mpc}^{-1}]$  \\
\hline
 Planck & $7.3 \times 10^{-11}$ & 0.0059 & 0.011 & 0.0015 \\ \hline \hline
+ LSST & $6.1 \times 10^{-11}$ & 0.0019 & 0.0023 & 0.00031 \\ \hline
 + Euclid & $3.2 \times 10^{-11}$ & 0.0031 & 0.0027 & 0.00041 \\ \hline
\end{tabular}
 \caption{\small ({\bf A full standard clock}) Marginalized $1-\sigma$ constraints on a full standard clock template. We have set $C=0.0307$ and $ k_r =0.109$ Mpc$^{-1}$. Each LSS survey is considered  in combination with ``Planck".   See also Fig. \ref{Fig:2d_SC} for a contour plot.}
\label{Table:SC}
  \end{center}
\end{table}

\begin{figure}
\center
\includegraphics[scale=.75]{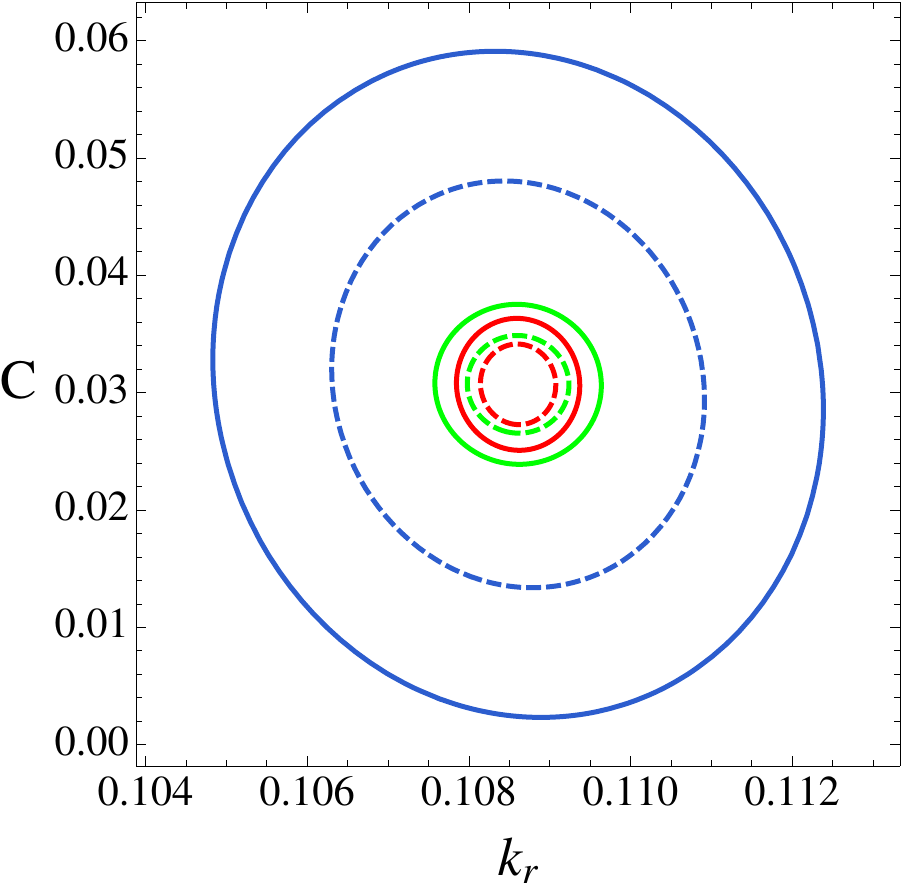}
\caption{\small ({\bf A full standard clock}) Contour plot for the full standard clock signal. The feature parameters are $C=0.0307$ and $ k_r =0.109$ Mpc$^{-1}$. The blue, red, and green contours are for Planck, Planck+LSST-like, and Planck+Euclid-like observations, respectively.}
\label{Fig:2d_SC}
\end{figure}

It is clear from these results that  LSS surveys will be able to greatly improve  constraints on both classes of feature models. Generally we expect improvements over the current Planck constraints by a factor of $\sim 2-10$, and details depend on the frequency  parameter of the model.
Because LSS surveys carry  the three-dimensional information in the sky, a clear advantage over CMB is its increased sensitivity for higher frequency models. As we can see from the tables, while the CMB generally gives poorer constraints for higher frequency models,\footnote{The sharp feature template with $k_f=0.1$ ${\rm Mpc}^{-1}$ is an exception, because the frequency is too low and this template does not oscillate much in the observable range.} LSS surveys give equally good constraints for models with very different frequencies. Therefore, the higher frequency models investigated in this study benefit more from the LSS surveys.

In particular, two benchmark models here are chosen to approximately represent the (formal)  best-fit models in the Planck 2015 data: one with a  sharp feature (with $k_f\approx 0.004{\rm Mpc}^{-1}$) \cite{Chen:2014cwa},  and another with a  resonance feature (with $\Omega\approx 30$) \cite{Ade:2015lrj}. Both candidates pick up signals around the wiggles near $\ell \sim 800$ and are  statistically insignificant in current data. With  future LSS surveys, the error bar for these two models will  shrink by a factor of 6, confirming or ruling out these models with high statistical significance.

As an exception, the constraints on the  step model that is used to fit the low $\ell$ glitch, namely Table.~\ref{Table:step}, will not gain much improvement from the LSS surveys. The reason is simply that the template for the step model (with the current choice of parameters) has a strongly scale-dependent envelop, which rapidly kills the signal at scales relevant to the LSS observations. We stress, however, that there is still a reasonable region in parameter space of the step model where the template can be well approximated by the sharp feature template, where no envelop is imposed (see Fig. \ref{Fig:step}).

In Tables \ref{Table:alt_clock} and \ref{Table:inf_clock} we summarize the constraints for the clock signals of different primordial Universe scenarios (see also Figs. \ref{Fig:2d_clock_alt} and \ref{Fig:2d_inf_clock}). The marginalized 1-$\sigma$ constraints and the joint 2D contour plots for a full clock signal template are also presented in Table \ref{Table:SC} and Fig. \ref{Fig:2d_SC}.

Compared to the sharp feature and resonance feature templates, the clock signals have one or two more free parameters. As a result, the errors of the amplitudes $C$ slightly increase.

Comparing the LSS with the Planck only results, we again see an overall improvement by a factor of $5$ to $10$. Also in this case, the two examples considered here, namely in Table.~\ref{Table:inf_clock} and \ref{Table:SC}, represent the current (not statistically significant) best-fit models, both fitting the wiggles in the CMB angular power spectrum around $\ell\sim 800$.  For these two candidate models, the error-bars on $C$ will be reduced by a  factor of 5, demonstrating the power of the future LSS surveys.
In particular note that if $C$ is larger than 0.01, these surveys will be able to discriminate between inflation and alternatives (e.g., contraction or slow expansion) by constraining the value of  the parameter $p$. While there is not a theoretically-motivated prior for $C$ and it could be vanishingly small, this offers nevertheless an unprecedented window into the very early Universe.

As for the bump feature, we have summarized the expected constraints in Table \ref{Table:local}. The improvement from LSS surveys over the current Planck constraint are not as impressive as in previous cases. This is because these signals do not oscillate as a function of scale, so the advantage of 3D resolution is not as significant as in the oscillatory case,  reducing the  impact of LSS observations on the constraints.

\begin{table}[t]
  \begin{center}
\begin{tabular}{|c||c|c|c|c|c|}
\multicolumn{1}{l}{}\\
\multicolumn{3}{r}{$k_f=0.05/{\rm Mpc}   \hspace{-.2cm}$}
\\
\cline{2-5}
\multicolumn{1}{c|}{}
&$A_s$ & $n_s$ & $C$ &$ k_f[{\rm Mpc}^{-1}]$   \\
\hline
 Planck & $7.4 \times 10^{-11}$ & 0.0065 & 0.0071 & 0.013 \\ \hline \hline
 + LSST & $6.3 \times 10^{-11}$ & 0.0024 & 0.0031 & 0.0051 \\ \hline
 + Euclid & $3.3 \times 10^{-11}$ & 0.0037 & 0.0040 & 0.0069 \\ \hline
\multicolumn{1}{l}{}\\
\multicolumn{3}{r}{$k_f=0.1/{\rm Mpc}   \hspace{-.2cm}$}
\\
\cline{2-5}
\multicolumn{1}{c|}{}
&$A_s$ & $n_s$ & $C$ &$ k_f[{\rm Mpc}^{-1}]$   \\
\hline
 Planck & $7.7 \times 10^{-11}$ & 0.0091 & 0.012 & 0.031 \\ \hline \hline
 + LSST & $6.4 \times 10^{-11}$ & 0.0032 & 0.0044 & 0.012 \\ \hline
 + Euclid & $3.3 \times 10^{-11}$ & 0.0062 &0.0087 & 0.020 \\ \hline
\multicolumn{1}{l}{}\\
\multicolumn{3}{r}{$k_f=0.2/{\rm Mpc}   \hspace{-.2cm}$}
\\
\cline{2-5}
\multicolumn{1}{c|}{}
&$A_s$ & $n_s$ & $C$ &$ k_f[{\rm Mpc}^{-1}]$   \\
\hline
 Planck & $7.9 \times 10^{-11}$ & 0.0099 & 0.031 & 0.34 \\ \hline \hline
+ LSST & $6.5 \times 10^{-11}$ & 0.0038 & 0.0069 & 0.085 \\ \hline
+ Euclid & $3.2 \times 10^{-11}$ & 0.0059 & 0.027 & 0.30 \\ \hline
\end{tabular}
 \caption{\small ({\bf Bump features}) Marginalized  1-$\sigma$ constraints on the free parameters for the localized feature model at different locations (different $k_f$). The amplitude has been set to $C=0.01$. }
\label{Table:local}
  \end{center}
\end{table}

It is also interesting to compare the constraints from the LSST-like experiment with the Euclid-like experiment.

As we can see, the Euclid-like experiment provides better constraints on the amplitude of the matter  power spectrum $A_s$.
In the absence of  the redshift-space distortion signal (i.e. $\beta=0$ in Eq. \eqref{Eq:Pg}), the amplitude of the power spectrum is totally degenerate with the bias parameter. The degeneracy is, however, broken by the redshift-space distortions. Spectroscopy has better redshift resolution, which yields a better measurement of the redshift space distortion signal, and hence better measurement of $A_s$.

On the other hand, we can see that the  LSST-like experiment provides better constraints on the other parameters, including the feature model parameters. Such an experiment surveys a larger volume with higher galaxies density, increasing the number and precision of available modes in the analysis. It turns out that, for the purpose of constraining the scale-dependent oscillatory signals, this advantage slightly wins over the disadvantage in the redshift measurement.

\section{Conclusions and outlook}
\label{sec:conclusions}

Primordial features are one of the most important extensions to the standard $\Lambda$CDM  cosmological model, carrying important  information about high energy physics and the early stages of the Universe. The signatures of new physics, discrimination of the early Universe scenarios, discovery of massive particles during inflation, and detection of fine structures in the inflaton potential are a sample of different motivations for studying  primordial features, showing their significant potential impact in our understanding of the Universe. In this work we have shown that large-scale structure observations provide an impressive opportunity for primordial features discovery.  A broad class of primordial features are studied using the Fisher matrix approach to forecast the detectability and constraining power of forthcoming observations. These observations include the most ambitious ground-based photometric survey (LSST) and space-based  spectroscopic mission (Euclid), both of which are expected to take data in the next decade. We compare the performance of these constraints with the current constraints from Planck 2015 results.

Here is a summary of the main results.
\begin{itemize}

\item We have classified and studied the following several classes of models: 1) Sharp feature models, in which the inflation model contains a sharp feature or an injection of new physics at a specific moment; 2) Resonance feature models, in which the inflation model contains features that are periodic in time or an injection of new physics at a specific energy scale; 3) Standard Clocks models, in which the oscillation of massive fields is used as a clock to measure the background scale factor evolution $a(t)$, which can in principle distinguish inflation from alternative scenarios; and 4) Bump feature models, in which some large interactions in the model create a bump in the power spectrum around a certain moment.

\item Compared to the CMB, a significant advantage of LSS surveys is their resolution on the scale-dependent oscillatory signals. In the CMB, such signals are smoothed out in the 3D to 2D projection, while LSS preserves at least part of the 3D information. As a result, LSS surveys are sensitive to models with a much broader frequency range. Some high-frequency models, that are poorly constrained by CMB, are expected to be much better constrained with LSS surveys.

\item Depending on the frequency considered, the amplitude of most models studied in this paper is expected to be better constrained by a factor of 2-10 by LSS surveys. In particular, we studied several benchmark examples which represent (formal) best-fits models found in the Planck 2013/2015 data. These are interesting candidates for new physics, but their statistical significance is low or marginal. We show that the LSS surveys will be able to reduce the error bars by a factor of approximately 5, and hence reach definite conclusions on these candidates in the next decade or two.

\item Constraints on models whose signals are restricted only in the largest scales, such as the step model used to explain the $\ell\sim 20-40$ glitch, will not be significantly improved, because the signal has already decayed at scales relevant to LSS. Constraints on models whose signals do not oscillate in scales, such as the bump feature, will also be improved but less than the oscillatory ones, because the constraint from CMB is also competitive.

\item Both Euclid-like and LSST-like experiments are expected to considerably advance our knowledge and  improve the constraints in the feature models. The Euclid-like survey typically provides a better constraint on the amplitude of the leading order primordial power spectrum due to its higher resolution in redshift measurement; while the LSST-like survey provides better constraints on other parameters such as the feature model parameters, due to its larger volume coverage and lower shot noise (more measured samples per volume).

\item Finally, we note that the constraints on the amplitude $C$ of the features, which is the most relevant  quantity  for  feature discovery, is almost independent of its assumed fiducial value as long as $C$ is small. This makes our main qualitative conclusion robust and generalizable to other fiducial amplitudes.

\end{itemize}

Let us end by briefly sketching the outlook for future studies.

In this work we have focused on the LSS power spectrum. Feature models generically predict associated signals in higher-point correlation functions \cite{Chen:2006xjb,Chen:2008wn,Adshead:2011jq,Gong:2014spa,Mooij:2015cxa}.
These non-Gaussianities have highly correlated frequencies determined by the power spectrum, but with model-dependent amplitudes. Therefore, although the observable non-Gaussianities are not guaranteed, observing them with the right frequencies would provide extremely important supportive evidence for feature detection and information for model selection. See Refs. \cite{Fergusson:2014hya,Fergusson:2014tza,Munchmeyer:2014cca,Meerburg:2015owa,Meerburg:2015yka} for advances in this direction in the CMB. It would be interesting to forecast how LSS surveys would be able to contribute to this.

We have only analyzed  the linear matter power spectrum in this work. Non-linear corrections to the power spectrum, despite the fact that  they introduce extra uncertainties, can help to improve the statistical significance of feature models. In particular, it is interesting to note that the non-linear evolution of structures  does not generate patterns similar to those from oscillatory features. Although the amplitudes of the features can be washed out by  mode-coupling introduced by non-linearities in a similar way as the baryon acoustic oscillation feature is suppressed, some signal remains even in the mildly non-linear regime (e.g., see \cite{Vlah:2015zda} for a recent application to primordial wiggles in the power spectrum).

Another very powerful probe of primordial features is the CMB polarization power spectrum.
Polarization predictions can be used to move beyond  posterior inferences that currently dominate the field. Primordial features should also have a signature in the CMB E-mode polarization, and they present a higher significance than that coming from temperature alone \cite{Dvorkin:2007jp}. In particular, the signature in the CMB E-mode polarization from an inflationary model with a step in the potential constructed to  match the  features seen in the temperature power spectrum at multipoles of $\ell=20-40$,
should have the power of confirming or ruling out their primordial origin at the $\sim 3 \sigma$ significance with  nominal Planck polarization sensitivity  \cite{Mortonson:2009qv,Dvorkin:2011ui,Miranda:2014fwa},  and at an even higher significance with a cosmic variance limited experiment.
LSS are likely not to improve much  CMB constraints  on features whose effect in the power spectrum is concentrated at low $\ell$ ($< 100$). However, for  features  which affect the power spectrum over ranges of scales well probed by LSS, these surveys offer a unique probe.
The improvement  achievable from  polarization is likely to lie between the temperature and LSS data. Details should depend on models and frequencies.
More systematic forecast on constraints from CMB polarization on feature models is left for a future study.

Because there is no theoretical lower-limit for the amplitudes of primordial features, it is important to know the limit of the experimental reach. This would be set by the 21 cm tomography, at least in principle, which is explored in \cite{feature21cm}.

Our results indicate  that the study of primordial features is likely to  receive a significant boost with the next generation of large-scale structure surveys, improving  quantitatively and qualitatively our knowledge of beyond standard-cosmological model physics and the physics behind inflation.

\acknowledgments
We thank Daniel Eisenstein, Avi Loeb, Daan Meerburg and Moritz Munchmeyer for fruitful discussions.
We thank Gabriel Perez and the system administration of the Hipatia computer.
XC is supported in part by the NSF grant PHY-1417421.
LV acknowledge support of Spanish MINECO grant AYA2014-58747-P and  MDM-2014-0369 of ICCUB (Unidad de Excelencia `Maria de Maeztu').


{}

\end{document}